\documentclass[pra,aps,amsmath,amssymb,superscriptaddress,longbibliography,notitlepage,twocolumn]{revtex4-1}
\usepackage{graphicx,multirow,outlines,ulem}
\usepackage{color}
\usepackage{hhline}
\usepackage{physics}
\usepackage{hyperref}
\usepackage{bbold}

\usepackage[mathscr]{eucal}

\usepackage{tikz}
\usepackage{xcolor}

\newcommand{\bo}{\begin{outline}}
\newcommand{\eo}{\end{outline}}

\newcommand{\qed}{\nobreak \ifvmode \relax \else
      \ifdim\lastskip<1.5em \hskip-\lastskip
      \hskip1.5em plus0em minus0.5em \fi \nobreak
      \vrule height0.75em width0.5em depth0.25em\fi}

\begin{document}

\title{Quantum-to-semiclassical Husimi dynamics of non-Hermitian localization transitions}
\author{Pallabi Chatterjee}
\email{ph22d001@iittp.ac.in } 
\affiliation{Indian Institute of Technology Tirupati, Tirupati, India~517619} 
\author{Bhabani Prasad Mandal}
\email{bhabani@bhu.ac.in }
\affiliation{Department of Physics, Banaras Hindu University, Varanasi~221005, India}
\author{Ranjan Modak}
\email{ranjan@iittp.ac.in}
\affiliation{Indian Institute of Technology Tirupati, Tirupati, India~517619} 
\begin{abstract}

The localization transition in the Hermitian Aubry--André model is known to have a classical origin, with the critical point being exactly predictable from an analysis of classical phase-space trajectories. Motivated by this correspondence, we investigate whether a similar classical origin exists for localization transitions in non-Hermitian quasiperiodic Hamiltonians. Using semiclassical Husimi dynamics together with a detailed phase-space stability analysis, we show that localization transitions persist even in the semiclassical limit of such non-Hermitian models. However, in sharp contrast to the Hermitian Aubry--André case, the transition point inferred from classical phase-space analysis does not coincide with the quantum critical point. Instead, we find that the semiclassical transition depends sensitively on the choice of the irrational parameter defining the quasiperiodic potential, indicating the absence of a universal classical–quantum correspondence for the localization transition in the non-Hermitian setting. Nonetheless, we identify a suitable parameter regime in which the classical dynamics can faithfully mimic the quantum dynamics over a finite but appreciable time window.

%We study localization–delocalization transitions in quasiperiodic lattice systems using phase-space Husimi dynamics. Comparing lattice transport with the quantum and semiclassical Husimi evolution of corresponding continuous models, we find that while the quantum Husimi distribution faithfully captures the localization transition, trajectory-based semiclassical Husimi dynamics fail to reproduce the correct quantum critical point. In non-Hermitian quasiperiodic lattices, the semiclassical transition follows a $\beta$ dependent geometric criterion that generally underestimates the quantum transition. Benchmarking against Hermitian models shows that, although semiclassical dynamics succeeds for the Aubry–André model, it also fails for the generalized Aubry–André model by yielding a single energy-independent critical parameter and missing the quantum mobility-edge structure. These results reveal a fundamental limitation of semiclassical phase-space intuition in both non-Hermitian and mobility-edge quasiperiodic systems.
\end{abstract}
\maketitle

\section{Introduction}
In one-dimensional (1D) non-interacting systems, arbitrarily weak uncorrelated disorder leads to Anderson localization of all single-particle states, resulting in the complete absence of transport \cite{anderson.1958,abrahams1979scaling}. In contrast, quasiperiodic lattice models of non-interacting particles offer a clean and controllable framework for exploring localization phenomena beyond the traditional Anderson paradigm, which can even support localization-delocalization transition in 1D~\cite{aubry.1980}. A quasiperiodic potential can be defined as a deterministic potential characterized by an incommensurability parameter $\beta$, such that when $\beta$ is irrational with respect to the underlying lattice spacing, the potential remains quasiperiodic. With recent experimental advances, particularly in cold atoms and photonic systems, the localization–delocalization transition in these models can now be probed with remarkable precision \cite{exp1,exp2,exp3,exp4}. A wide variety of quasiperiodic potentials have been investigated, revealing rich phase diagrams and diverse localization properties \cite{aa_addi1,aa_addi2,aa_addi3,aa_addi4,aa_addi5, AA_1}.

Extending these ideas to open systems \cite{open_1,open_2,open_3,open_4, open_PT}, non-Hermitian Hamiltonians have emerged as a powerful framework for describing non-interacting systems in which energy or particles can exchange with the environment, thus going beyond the conventional Hermitian paradigm. In such non-Hermitian settings, disorder can generate unconventional localization phenomena and metal–insulator transitions. For instance, in a non-Hermitian Anderson-type model with purely imaginary on-site disorder, states can become exponentially localized, showing that non-Hermiticity promotes localization~\cite{nh_1}. In contrast, introducing asymmetry in the hopping amplitudes—as in the Hatano–Nelson model \cite{Longhi_2023}—produces a delocalization transition, illustrating the competition between non-reciprocal transport and disorder.
Non-Hermitian extensions of different variants of quasiperiodic models can exhibit even richer behavior. These systems can feature modified localization–delocalization transitions, mobility edges, and entirely new spectral and topological phase transitions that are absent in Hermitian analogs \cite{nh_2, nh_3, nH_6, nH_7,nh_qp_1,nh_qp_2,nh_qp_3,nh_qp_4,nh_qp_5,nh_qp_6,nh_qp_7,nh_qp_8,nh_qp_9}. By tuning the complex phase or amplitude of the quasiperiodic potential, it is possible to manipulate the interplay between conventional Anderson localization and the non-Hermitian skin effect \cite{nh_4}. Experimental simulation demonstrated how non-Hermitian effects, combined with disorder, can generate both localization and topologically nontrivial states~\cite{nh_5}. 
In general, non-Hermitian systems have attracted growing interest \cite{non_hermitian_topology}, motivated by experimental realizations in photonic lattices and ultracold atomic gases~\cite{nh_exp_1,nh_exp_2,nh_exp_3,nh_exp_4,nh_exp_5}. A celebrated paradigm in non-Hermitian physics is the parity–time (PT) symmetric class, namely a broad family of non-Hermitian Hamiltonians first introduced by Bender et al., which can exhibit entirely real spectra as long as  the PT symmetry is unbroken~\cite{bender_98,bender2007making}. PT-symmetric quasiperiodic lattice models have also been realized and have attracted considerable attention~\cite{nh_pt_1,nh_pt_2}.
Taken together, these developments establish non-Hermitian quasiperiodic lattices as a powerful platform for investigating unconventional localization phenomena, spectral topology, and the role of openness in quantum systems.

An intriguing question, however, is whether the delocalization–localization transitions observed in quantum quasiperiodic models persist in an analogous classical description. Earlier studies have shown that for the simplest quasiperiodic model—the Aubry–André (AA) model \cite{aubry.1980}—the quantum critical point can be accurately reproduced from the corresponding classical trajectory \cite{semiclassical_1}. In this work, we extend these ideas to non-Hermitian quasiperiodic systems and investigate whether a classical description can capture such localization–delocalization transitions.

%we further analyze additional quasiperiodic models that are known to exhibit mobility edges, and examine whether their classical trajectories capture the same transition behavior. 

There are several established approaches for the classical-quantum correspondence \cite{kick_top_open} and treating non-Hermitian systems in phase space. These include formulations based on position and momentum using complexified classical equations of motion with complex phase-space variables \cite{NH_phase_space_1}, as well as descriptions on a real phase space with a modified metric structure, particularly within Gaussian-state dynamics \cite{NH_phase_space_2}. Related work has derived canonical equations of motion using coherent-state approximations \cite{Non_hermitian_traj}, and applied phase-space equations obtained from Gaussian wavepacket approximations to single-band non-Hermitian tight-binding chains \cite{NH_phase_space_3}. As well as there is a study of complexified semiclassical theory \cite{complex_semiclassical}. More recently, it has been shown that for non-Hermitian Hamiltonians, a semiclassical approximation leads to an evolution equation for the Husimi distribution~\cite{Semiclassical_prl_1,semi_classical_prl_2}. The Husimi distribution is strictly positive and normalizable, thereby providing a well-defined phase-space representation of quantum dynamics \cite{Husimi_1,Husimi_2,Husimi_3,Husimi_4,Semiclassical_prl_1,semi_classical_prl_2,NH_phase_space_4}. For the quasiperiodic model, it provides valuable phase-space intution about delocalization-localization transition \cite{Phase_AA_eigen}. For continuous models, the semiclassical Husimi evolution follows canonical phase-space trajectories \cite{Graefe_2010}, multiplied by a norm landscape \cite{Semiclassical_prl_1}, which reduces to unity in Hermitian systems. It has further been established that quantum and classical Husimi flows coincide exactly when the Hamiltonian is bilinear in the creation and annihilation operators, whereas for more general Hamiltonians, the classical Husimi dynamics constitute only a short-time approximation to the full quantum evolution. This motivates us to adopt the Husimi approach and examine whether the time evolution of the semiclassical Husimi function can accurately predict the quantum localization–delocalization transition point in non-Hermitian quasiperiodic  systems \cite{non_hermitian_longhi_1,non_hermitian_longhi_2}. 

%We compute the time evolution of the Husimi distribution for an initially coherent state governed by the continuous version of the Hamiltonian and compare its propagation speed with that obtained from the full lattice dynamics, as well as with the semiclassical limit of the corresponding quantum Husimi evolution.

Our results show that, in contrast to the Hermitian AA  model, the classical Husimi phase-space dynamics of the non-Hermitian model fail to reproduce the correct quantum transition point. This breakdown highlights a fundamental limitation of trajectory-based semiclassical descriptions in non-Hermitian quasiperiodic systems. We further determine the transition point from a fixed-point analysis of the classical trajectory geometry and show that the classical Husimi dynamics follow this geometric transition. Moreover, unlike the Hermitian AA model, the transition point in the non-Hermitian lattice depends explicitly on the incommensurability parameter $\beta$, reflecting the irrational quasiperiodicity of the potential. Notably, we identify a special value of $\beta$ for which the classical and quantum transition points coincide. 

The manuscript is organized as follows. In Sec.~\ref{secII}, we describe the lattice models, and in Sec.~\ref{secIII} we discuss the corresponding lattice dynamics. Section~\ref{secIV} is devoted to the quantum Husimi dynamics of the continuum models. In Sec.~\ref{secV}, we present the semiclassical limit of the models and the associated semiclassical Husimi dynamics. \textcolor{black}{In section \ref{Dephasing} we discuss dephased dynamics and compare it with the actual quantum dynamics.} Finally, we conclude in Sec.~\ref{secVI}.

\section{Model}\label{secII}
We analyze two non-Hermitian models, referred to as Model I and Model II. In Model I, Hermiticity is broken through asymmetric hopping, whereas in Model II, non-Hermiticity arises from a complex on-site potential.
\subsection{Model I}

Model I is the non-Hermitian version of the AA model, where the non-Hermiticity has been introduced by the unequal left ($J_L$) and right ($J_R$) hopping terms, and is described by the following Hamiltonian on a lattice of size $L$,
\begin{equation}
    H_I=\sum_j( J_L c_j^\dagger c_{j+1}+ J_R c_{j+1}^\dagger c_j ) + 2V\sum_j  \cos(2\pi \beta j)c_j^{\dagger}c_j,
    \label{Model_1_lattice}
\end{equation}\\
here, $c_j^\dagger, c_j$ are the creation and annihilation operators respectively, and $V$ is the strength of the quasiperiodic potential. $\beta$ is an irrational number. In the Hermitian case, there is a delocalization-localization transition as one tunes the parameter $V$, and the transition point is at $V=J_L=J_R$~\cite{aubry.1980}. On the other hand, this non-Hermitian model 
is well studied in ~\cite{Longhi_2023,tanmoy_dna}, where it is found that the transition point of delocalized to localized phase is at $V_c=J_L$ with $J_L>J_R$. In most of our calculations, we set $ J_L=1$ and $ J_R=1/2$.
%The speed of the excitation transport $v$ decreases continuously with $V$ and goes to zero at the localized phase. Unlike the Hermitian case with $J_L=J_R$, in the case of a non-Hermitian case, it is found that the phase transition is discontinuous in terms of the speed of the excitation transport $v$, which suggests that, very close to the critical $V_c$, there could be a possible finite ballistic speed before it goes to zero as we increase $V$. 

\subsection{Model II}
Another model we study features non-Hermiticity originating from the on-site potential, with the Hamiltonian given by~\cite{non_hermitian_longhi_1},
\begin{equation}
    H_{II}=J\sum_j(  c_j^\dagger c_{j+1}+  c_{j+1}^\dagger c_j ) + V\sum_j  e^{(-2\pi i \beta j)}c_j^{\dagger}c_j.
    \label{Model_2_lattice}
   \end{equation}
   Previous numerical results for model II suggest a metal-insulator phase transition at the critical value $ V_c=J$, which exhibits a PT-symmetric breaking phase transition as well. For $V_c<J,$ the energy spectrum is entirely real, and all eigenstates are delocalized (metallic and unbroken
PT phases), while for $V_c> J$ the energy spectrum becomes
complex and all eigenstates are localized (insulating and PT
broken phases). In all our calculations, we set $J=1$. For most of the data presented in this manuscript, we choose $\beta = (\sqrt{5}-1)/2$; scenarios with different values of $\beta$ are discussed only in Sec.~\ref{secV}.

\begin{figure}[t]
    \centering
    \includegraphics[width=0.45\textwidth]{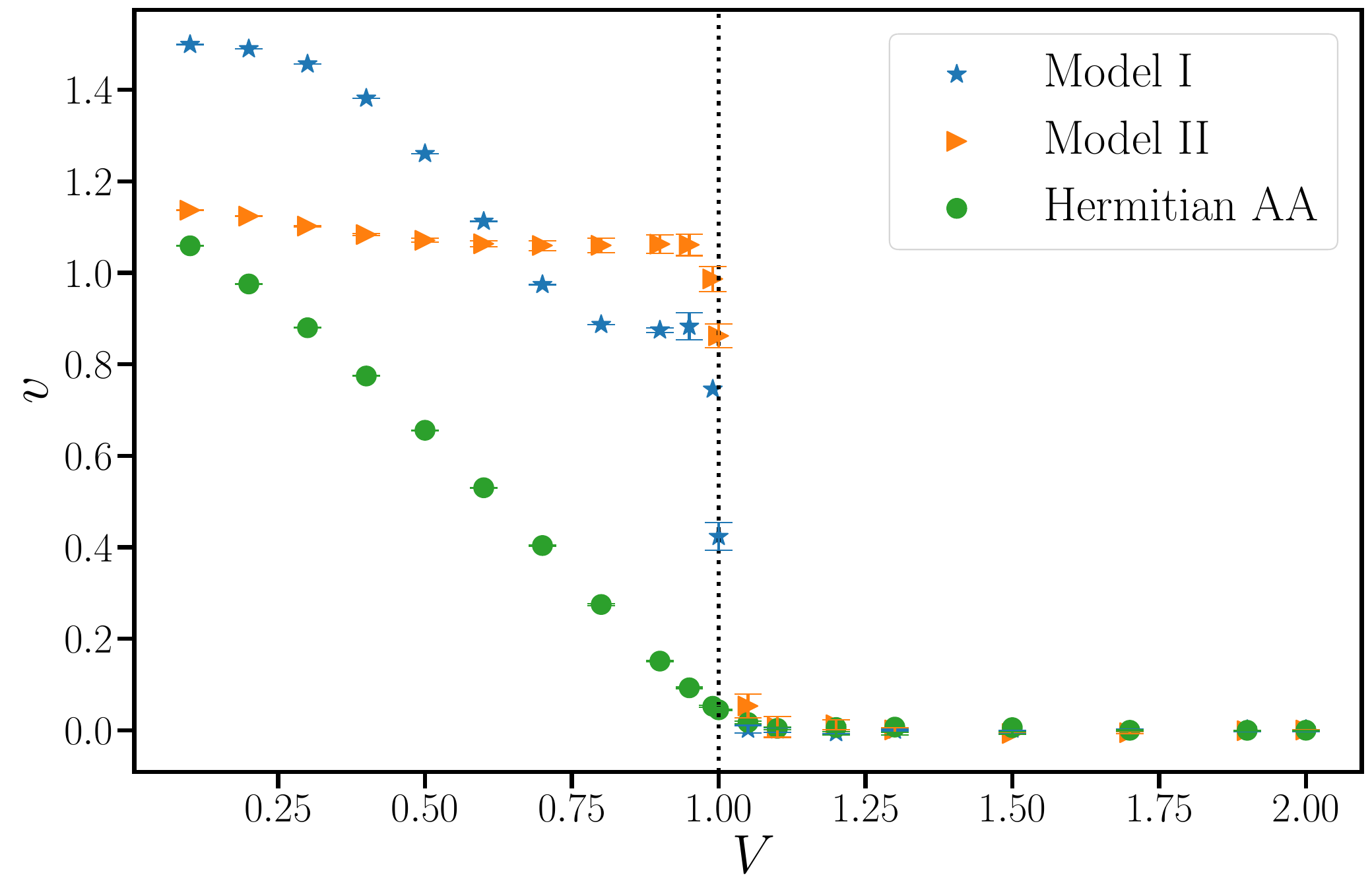}
    \caption{Velocity of the excitation propagation $v$ vs. $V$. Results are for Model I, Model II, and the Hermitian AA model. Results are with a starting coherent initial state centred at the centre of the lattice. The black dotted line represents the delocalization-localization transition point. %System size is taken as $L=601$.
  \textcolor{black}{Results shown here are extrapolated thermodynamic findings}.}
    \label{dy_lattice}
\end{figure}

\begin{figure*}
    \centering
    \includegraphics[width=0.85\textwidth, height=0.65\textwidth]{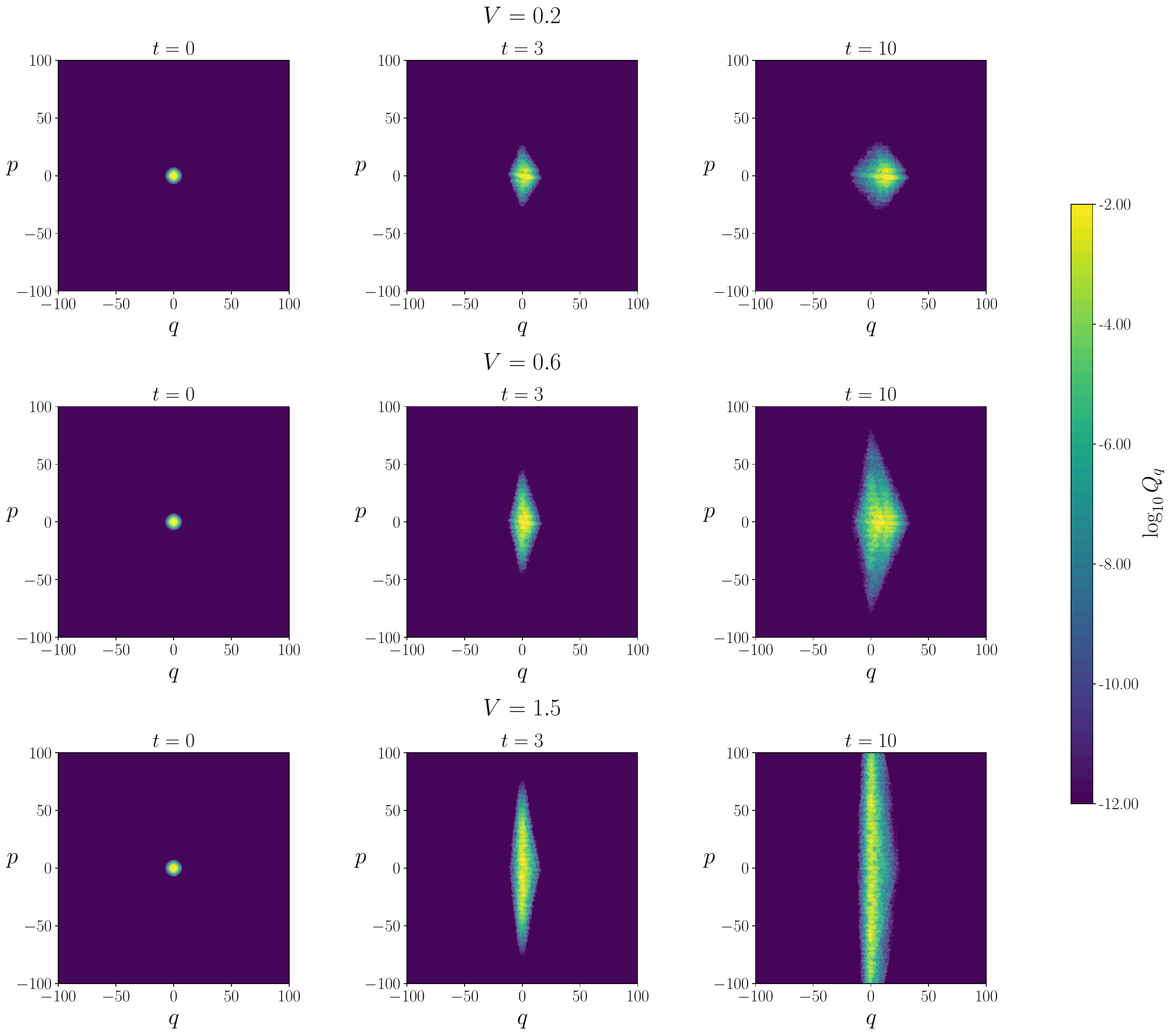}
   
    \caption{Quantum Husimi distribution at different time instances starting from coherent state $|z=0\rangle$ and evolving under the Hamiltonian [Eq. (\ref{continuous_model_I}) ] continuous version of the Model I. Data is for $V=0.2, 0.6,1.5.$ The color code represents the Husimi values. }
    \label{Q_1}
\end{figure*}

\begin{figure*}
    \centering
     \includegraphics[width=0.85\textwidth, height=0.65\textwidth]{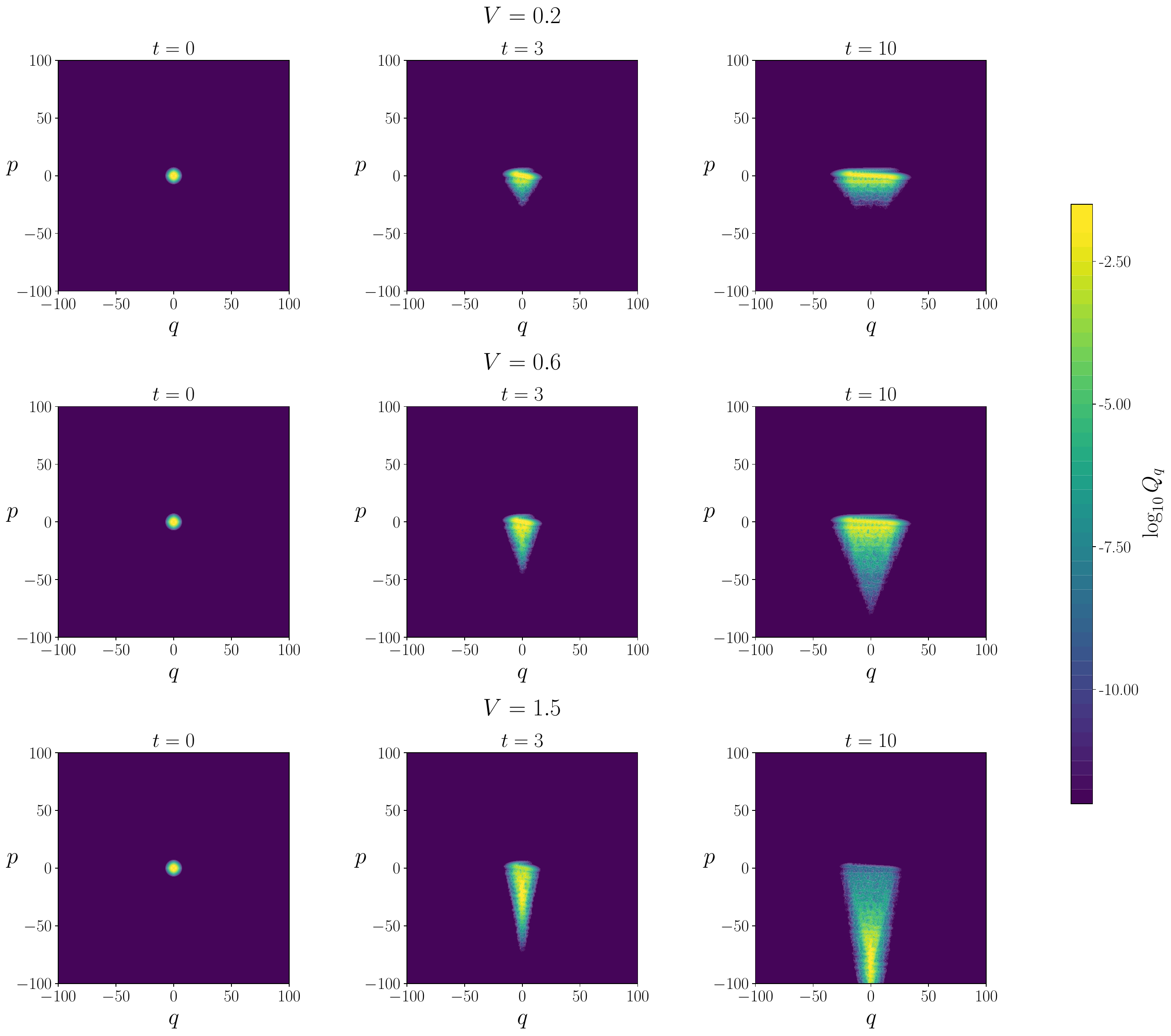}

    \caption{Quantum Husimi distribution at different time instances starting from coherent state $|z=0\rangle$ and evolving under the Hamiltonian [ Eq. (\ref{continuous_model_II})] continuous version of the Model II. Data is for $V=0.2, 0.6, 1.5$. The color code represents the Husimi values. }
    \label{Q2_1}
\end{figure*}
\begin{figure}
    \centering
    \includegraphics[width=0.45\textwidth]{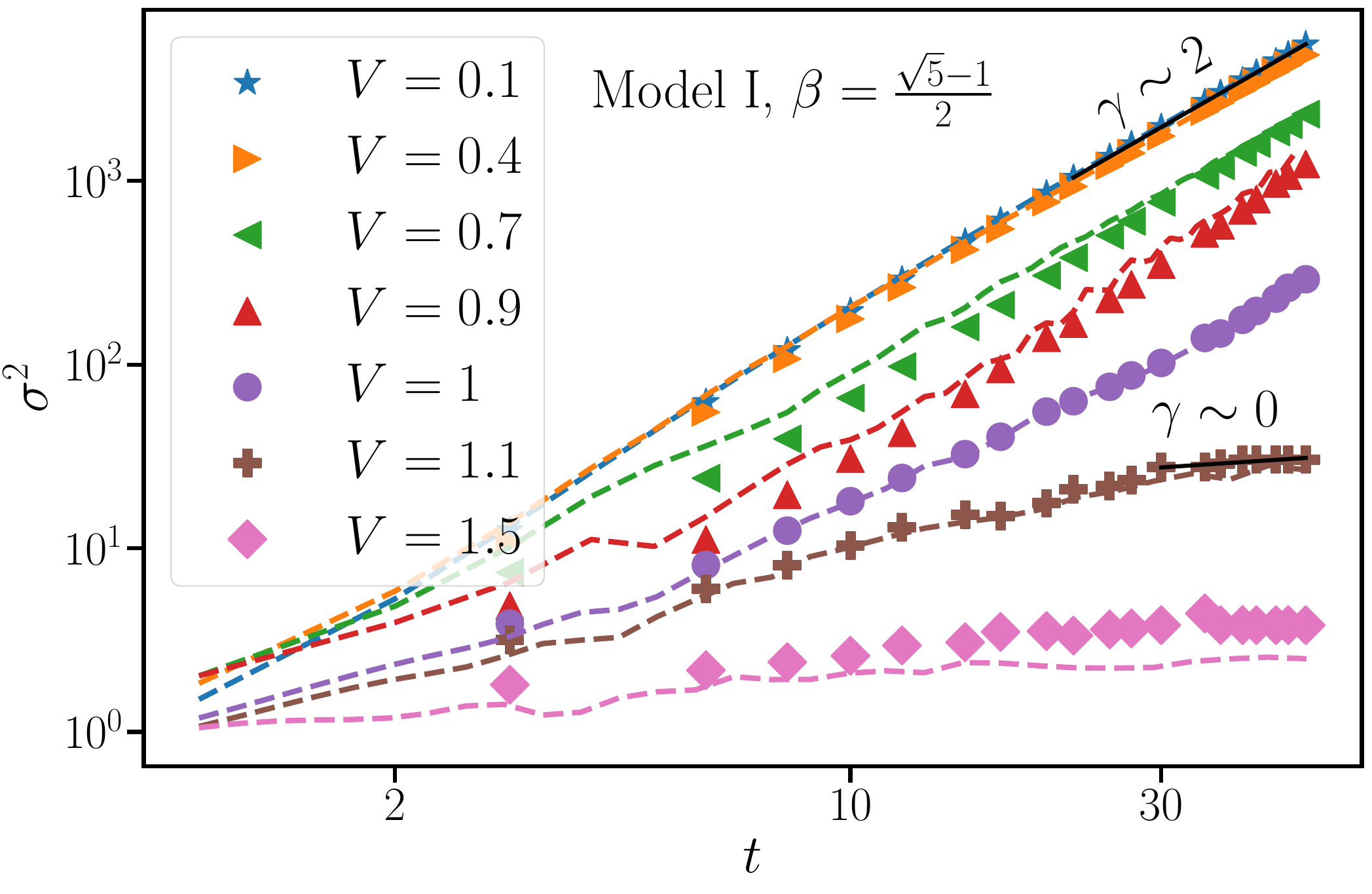}
    \includegraphics[width=0.45\textwidth]{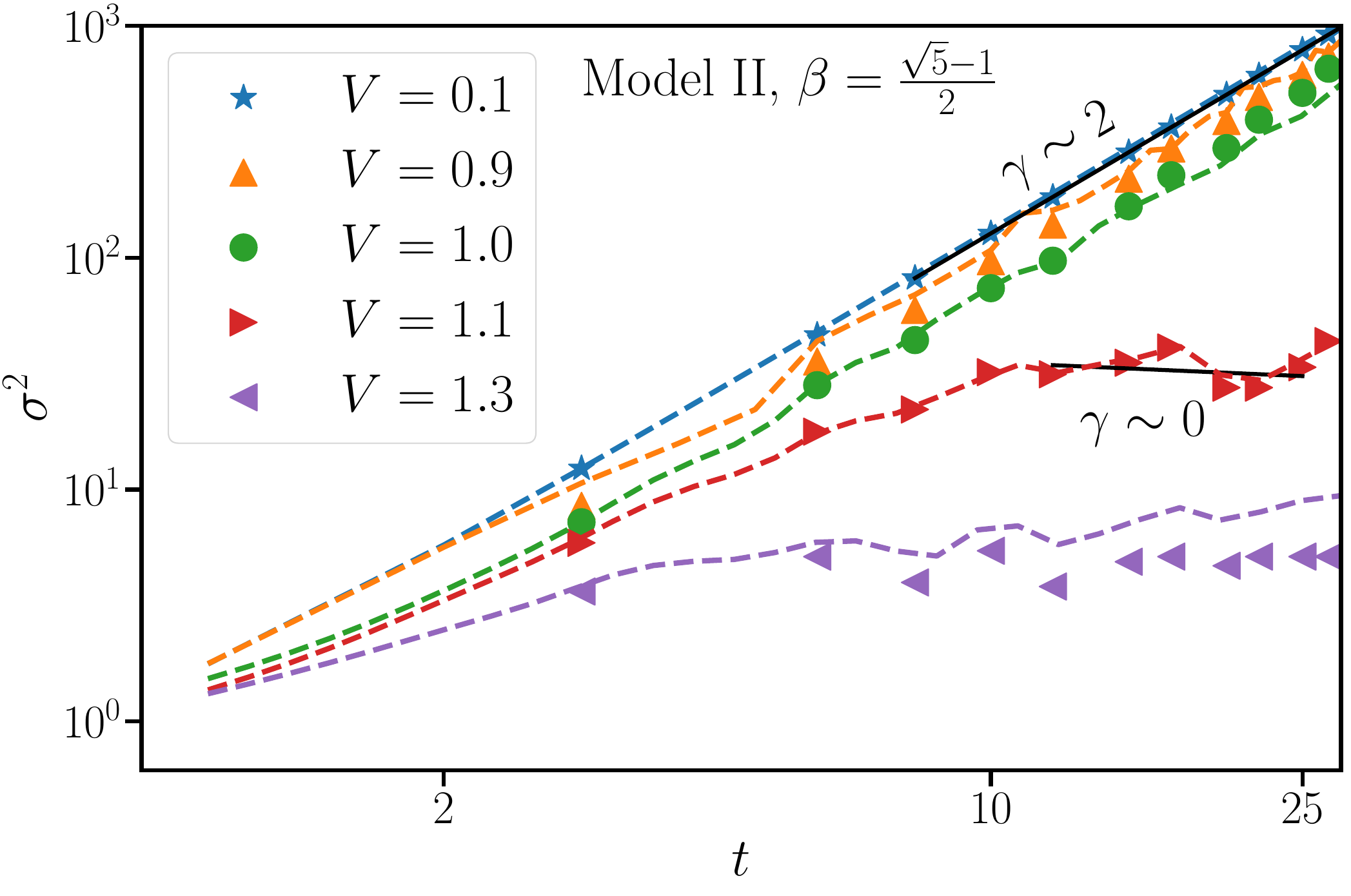}
      \caption{Upper panel: Comparison of $\sigma^2  $ vs. $t$ plots for non-Hermitian quantum continuous  Model I [Eq. (\ref{continuous_model_I}) ] and its corresponding lattice model, for $\beta=\frac{\sqrt{5}-1}{2}$. The symbols represent the quantum continuous model, and dashed lines represent the data for its lattice version. The lower panel represents the same for Model II\eqref{continuous_model_II} and its lattice version. The black solid line represents fitting to a function $\sigma^2 \sim t^{\gamma}$.}
   \label{comp_lattice_cont}
\end{figure}

\section{Lattice dynamics}\label{secIII}
Figure \ref{dy_lattice} shows the dynamical properties of the non-Hermitian lattice models \ref{Model_1_lattice} and \ref{Model_2_lattice}, together with the Hermitian version of Model I, starting from a coherent state. We choose as the initial state a minimal-uncertainty (coherent) wave packet centered at the middle of the lattice. Its position-space representation is
\begin{equation*}
\psi_j(t=0)
= \mathcal{N}\,
\exp\!\left(
-\frac{\bigl(j-\tfrac{L+1}{2}\bigr)^2}{2}
\right)
\end{equation*}
where 
$L$ is the system size, taken to be an odd integer, and 
$\mathcal{N}$ denotes the normalization constant.  Starting from this initial state, we calculate the velocity of excitation propagation, defined as
\begin{equation}
v = \lim_{t \to \infty} \lim_{L \to \infty} \frac{\sigma(t)}{t}.
\label{Eq: def_v}
\end{equation}
Here, \textcolor{black}{second order moment of the wave-packet spreading} is given by
\begin{equation}
\sigma^2(t) = \frac{\sum_j j^2 \lvert \psi_j(t) \rvert^2}{\sum_j \lvert \psi_j(t) \rvert^2}.
\label{Eq: sigma_lattice}
\end{equation}
The measure $v$ remains finite as long as the transport is ballistic, i.e., when $\sigma(t)\propto t$. In contrast, in the localized phase—and also in the diffusive phase, where $\sigma(t)\propto t^{1/2}$—this measure vanishes. For the non-Hermitian model studied here, the transport remains ballistic both in the delocalized phase and at the transition point (see Ref.~\cite{non_hermitian_longhi_2}). Consequently, the measure $v$ remains finite throughout this regime. In contrast, for the Hermitian version of Model~I, the dynamics at the transition point is diffusive (see Ref.~\cite{archak.2018}), leading to a vanishing value of $v$. Therefore, at the transition point, $v$ is expected to remain finite for the non-Hermitian model, in sharp contrast to its Hermitian model.

In Fig.~\ref{dy_lattice}, we present the dependence of the excitation transport velocity $v$ on the potential strength $V$. 
\textcolor{black}{
The data shown represent extrapolated thermodynamic-limit results. As evident from Eq.~\ref{Eq: sigma_lattice}, the determination of $v$ requires taking the appropriate long-time and thermodynamic limits. To this end, we perform an extrapolation of the finite-time data to obtain the asymptotic value of $v$. The extrapolation procedure is described in Appendix~\ref{extrapolated_results}.}.
As $V$ is increased, the velocity $v$ decreases and eventually vanishes upon entering the localized phase. While this behavior is continuous in the Hermitian case with $J_L = J_R$, the situation is qualitatively different for the non-Hermitian models. In particular, the transition at the critical point $V_c = 1$ is discontinuous when characterized in terms of the transport velocity $v$, in agreement with previous findings \cite{non_hermitian_longhi_2}. This discontinuity implies that, even arbitrarily close to the critical point from the delocalized side, the system can sustain a finite ballistic transport velocity, which then drops abruptly to zero as the potential strength is increased beyond $V_c$. We also observe that, for certain choices of initial states, the velocity $v$ initially increases with $V$, which is rather counterintuitive and suggests that, in some cases, the strength of the quasiperiodic potential can enhance the speed of excitation transport.
For all these calculations, we choose $\beta=\frac{\sqrt{5}-1}{2}$ for computing $v$. We emphasize that the discontinuity in $v$ near $V=1$ becomes increasingly pronounced with increasing $t$, and correspondingly with increasing system size $L$, which corresponds to the thermodynamic limit.

\section{Quantum Husimi Distribution of the Continuum Model}\label{secIV}
To gain a phase-space perspective on the underlying quantum dynamics using the Husimi distribution, we first construct the continuum version of the lattice model.
%Here, we study the dynamics of the Husimi distribution of the continuous version of the Model I~\eqref{Model_1_lattice} and Model II~\eqref{Model_2_lattice}. 
The fact that both models considered in this work can be diagonalized in the momentum basis for $V=0$ motivates our continuum formulation. Indeed, when $V=0$, both Model~I and Model~II are translationally invariant, and their Hamiltonians are diagonal in the momentum representation. In this representation, Model~I is described by
\begin{equation}
H_I (V=0) = \sum_p \left( J_R e^{ip} + J_L e^{-ip} \right) c_p^\dagger c_p ,
\end{equation}
while Model~II takes the form
\begin{equation}
H_{II} (V=0) = \sum_p J \cos p \, c_p^\dagger c_p .
\end{equation}
Here, $c_p^\dagger$ and $c_p$ are creation and annihilation operators in the momentum basis, and we set $\hbar=1$. 

Hence, we write the continuous version of Model I as,
\begin{equation}
    H_{I}^{q}=J_R e^{i\hat{p}} + J_L e^{-i\hat{p}} + 2V\cos(2\pi\beta \hat{q}),
    \label{continuous_model_I}
\end{equation}\\
with, $\hat{p}=\frac{i(\hat{a^\dagger}-\hat{a})}{\sqrt{2}}$ and $\hat{q}=\frac{(\hat{a^\dagger}+\hat{a})}{\sqrt{2}}$, where $a, a^\dagger$ are the Harmonic oscillator's (bosonic) creation and annihilation operators.

Similarly, for Model II, we write the continuous model as,
\begin{equation}
    H_{II}^{q}=2J \cos{\hat{p}} + V e^{(-2i\pi\beta \hat{q})}.
    \label{continuous_model_II}
\end{equation}
An analogous strategy has also been employed in earlier studies in different contexts~\cite{AA_semi_drive}.
%Again, in all our calculations, we take $J_L=1, J_R=0.5$ and $J=1$.  

Husimi distribution, which is a phase-space representation of the quantum state $|\psi(t)\rangle$, is given by~\cite{Semiclassical_prl_1, semi_classical_prl_2}.

\begin{equation}
    Q_q(t)=|\langle z|\psi(t)\rangle|^2,
    \label{quantum Husimi}
\end{equation}
where $|z\rangle= \hat{D}(z)|0\rangle=e^{z a^\dagger - z^* a}|0\rangle$, where $|0\rangle$ is the ground state of Harmonic oscillator and $z=(q + ip)/\sqrt{2}$ ($q$ and $p$ are eigenvalues of the position $\hat{q}$ and momentum $\hat{p}$ operators), hence $|z\rangle$ is a coherent state displaced by the displacement operator $\hat{D}$. 
We start from an initial \textcolor{black}{coherent state $|z=0\rangle$} and evolve it under the non-Hermitian Hamiltonian $H_{\mathrm{NH}}$ (which, in our case, corresponds to Eqs.~\eqref{continuous_model_I} and \eqref{continuous_model_II}). The time-evolved state is given by
$|\psi(t)\rangle = \frac{e^{-it H_{\mathrm{NH}} } |0\rangle}{\bigl\| e^{-it H_{\mathrm{NH}} } |0\rangle \bigr\|}$.
We then compute the overlap of this state with displaced coherent states at different points in phase space. The squared modulus of this overlap yields the quantum Husimi distribution at different time instants.
To quantify the spreading of the Husimi distribution with time, we also calculate the \textcolor{black}{second moment of the spreading of the Husimi distribution}  as,
\begin{equation}
    \sigma_H^2= \frac{ \int q^2 Q_q(q,p,t) dp dq}{ \int Q_q(q,p,t) dp dq}.
    \label{Eq: sugma_continuum}
\end{equation}
%\subsubsection{Quantum Husimi dynamics of Model I}
\paragraph*{Model I:} In Fig.~\ref{Q_1}, we show the time evolution of the Husimi distribution at different time instants for $V = 0.2$, $0.6$, and $1.5$ for the continuous version of Model~I [Eq.~\eqref{continuous_model_I}]. We observe that up to $V = 1$ the distribution continues to spread along the $q$ direction, whereas for $V > 1$ the spreading in the $q$ direction is suppressed. 
As is well known, wave-function propagation in the corresponding non-Hermitian lattice model is unidirectional~\cite{Longhi_2023}. The continuous quantum model faithfully reproduces this behavior and correctly captures the transition point at $V_c = J_L$.

We also present the time dependence of the \textcolor{black}{second moment} of the Husimi distribution in the upper panel of Fig.~\ref{comp_lattice_cont}. \textcolor{black}{The black solid line represents the fitting function $\sigma^2 \sim t^\gamma$}. In the small-$V$ limit, we observe $\sigma^2(t) \sim t^2$, indicating ballistic spreading. As $V$ is increased, the ballistic spreading gradually slows down, and for $V > 1$ the second moment essentially saturates, approaching $\sigma^2 \sim t^0$.  This behavior signals a delocalization--localization transition at $V_c = J_L$ for Model~I.
%{In our numerically accessible parameter regime, we simulate the dynamics up to time $t = 30$.} 
We also compare these results with the corresponding lattice calculations: the symbols represent data from the continuous model, while dashed lines denote the lattice dynamics.
\textcolor{black}{
To obtain smooth lattice data, we introduce a site-independent phase (flux) in the argument of the onsite cosine potential in $H_I$ and average the results over different values of this phase.}
Strictly speaking, Eq.~\eqref{Eq: sigma_lattice} and Eq.~\eqref{Eq: sugma_continuum} correspond to different quantities; therefore, their numerical values need not coincide. However, the qualitative time dependence of $\sigma^2(t)$ is very similar in both cases.

%Further, we compare the velocity of propagation of the wavepacket for the lattice model with the speed of propagation of the Husimi distribution of the continuous quantum model at a given $t=30$ for different $V$ values, followed by the equation,
%\begin{equation*}
 %   v=\sigma(t)/t.
%\end{equation*}
%In the Fig.~\ref{comp_actual_beta_model_1}, we see that the speed of propagation in the case of the lattice model and the quantum continuous model is very much qualitatively similar at the given time instant $t=30$, $\beta= (\sqrt{5}-1)/2$. 

%\subsubsection{Quantum Husimi dynamics of Model II}
\paragraph*{Model II} Similar to the previous model, we observe the time evolution of the quantum Husimi distribution for Model~II [Eq.~\eqref{continuous_model_II}] in Fig.~\ref{Q2_1} for different values of $V$. The spreading of the distribution in phase space correctly identifies the phase transition point at $V_c = J$: for $V \le V_c$ the distribution spreads along the $q$ direction, while for $V > V_c$ the spreading in the $q$ direction is suppressed. %All results shown correspond to $\beta = (\sqrt{5}-1)/2$ and $J = 1$.

We present the time dependence of the $\sigma^2$ of the Husimi distribution in the lower panel of Fig.~\ref{comp_lattice_cont}. We observe that $\sigma^2(t) \sim t^2$ for $V \le V_c = J = 1$, while it ceases to increase with time once $V > V_c$. This behavior indicates a delocalization--localization transition at $V_c = J = 1$. We also compare these results with those obtained from the corresponding lattice models and find reasonable good agreement between the two. 

In hindsight, these results confirm that the continuum version of our non-Hermitian lattice model provides a reasonably good approximation and efficiently reproduces the phase transition observed in the lattice models within the same parameter regime.

%Further, we compare the excitation propagation speed of the lattice model with the speed of propagation of the  Husimi distribution of the continuous quantum model at $t=30$. In the Fig.~\ref{comp_actual_beta}, we see that the variation of the speed of propagation with $V$ for the continuous model is very much similar to their corresponding lattice version at the given time instant $t=30$, $\beta= (\sqrt{5}-1)/2$, though they are not exactly on top of each other. 

%\begin{figure}[h]
%    \centering
    
   %  \includegraphics[width=0.48\textwidth]{quantum_sigma_2.pdf}
  %  \caption{$\sigma^2  $ vs. $t$ plot for non-Hermitian quantum for Model II\ref{continuous_model_II} for different $V$ values. Data is generated from the spread of the Quantum Husimi distribution in the phase space.}
%   \label{sigma_quantum}
%\end{figure}

\section{Semiclassical limit of the Husimi distribution}\label{secV}

In this section, we analyze the semiclassical limit of the quantum Husimi dynamics presented in the previous section. Our aim is to investigate whether the semiclassical description can capture and predict the localization–delocalization transition in the non-Hermitian Model~I and Model~II. 

\subsection{Formalism}
First, for the sake of completeness, we briefly outline the formalism required to take the appropriate semiclassical limit of the quantum Husimi dynamics, which has already been discussed in detail in Refs.~\cite{Semiclassical_prl_1, semi_classical_prl_2}.
The time evolution of the quantum Husimi function $Q_q$ (which was defined in  Eq.~\eqref{quantum Husimi}) under any non-Hermitian Hamiltonian $H_{\mathrm{NH}}$ could be written as,
\begin{align*}
    i \frac{\partial Q_q}{\partial t}= \langle z|H_{\mathrm{NH}}|\psi\rangle \langle \psi | z\rangle - \langle z |\psi\rangle \langle \psi |H_{\mathrm{NH}}^{\dagger}|z\rangle.
\end{align*}
Considering $H_{\mathrm{NH}}$ is an analytic function of creation and annihilation operators, in normal order form, we can expand any general non-Hermitian Hamiltonian as,
\begin{align*}
    H_{\mathrm{NH}}=\sum_{m,n=0}^{\infty} K_{m,n} a^{\dagger m}a^n, 
\end{align*}
where $K_{m,n}$ are complex coefficients. Then the evolution becomes,
\begin{align*}
    &i \frac{\partial Q_q}{\partial t}=\sum_{m,n} K_{m,n}z^{*m}\langle z | a^n |\psi\rangle \langle \psi |z\rangle \\
   & ~~~~~~~~~~~~~~~~~~~~~~~~~~-K_{m,n}^* z^m\langle z|\psi\rangle \langle \psi |a^{\dagger n}|z\rangle.
\end{align*}
One can use,
\begin{align*}
    &a^n|z\rangle=z^n|z\rangle, \\
    &\langle z |\psi\rangle\langle \psi | a^{\dagger n}|z\rangle=e^{-|z|^2}\frac{\partial^n}{\partial z^n}[Q_q(z)e^{|z|^2}].
\end{align*}
It is clear that, upon substituting these coherent-state identities, one obtains a higher-order differential equation for the Husimi distribution. If we neglect all higher-order derivatives beyond the first-order terms, the resulting equation for the evolution of the Husimi distribution reads as, 
\begin{align}
    \frac{\partial Q_{cl}}{\partial t} + i \frac{\partial H^c}{\partial z}\frac{\partial Q_{cl}}{\partial z^*}- i \frac{\partial H^{c*}}{\partial z^*}\frac{\partial Q_{cl}}{\partial z}-2\Gamma_I Q_{cl}=0,
    \label{semi_classical_dy}
\end{align}
which we refer to as the semiclassical evolution equation for the Husimi dynamics.
In Eq.~\eqref{semi_classical_dy}, we denote the Husimi distribution by $Q_{\mathrm{cl}}$ instead of $Q_q$ to emphasize that the solution of this equation represents the semiclassical approximation to the full quantum Husimi distribution $Q_q$.
Here $H^c$ is the classical counterpart of the non-Hermitian Hamiltonian, and can be written as,
\begin{align*}
    H^{c}=\sum_{m,n=0}^{\infty} K_{m,n} z^{*m}z^n,
\end{align*}
and the imaginary part of  $H^{c}$ is identified as $\Gamma_I=\frac{1}{2i}(H^{c}-H^{c*})$, and the real part as $\Gamma_R=\frac{1}{2}(H^{c}+H^{c*})$ . The semiclassical Husimi dynamics equation Eq.~\eqref{semi_classical_dy} can be solved using the method of characteristics, and one can find the solution as,
\begin{align}
    Q_{cl}(z,t)=Q_{cl}^0(\zeta_0(z,t))w(z,t).
    \label{semi_classical_soln}
\end{align}
This suggests starting from a given initial Husimi function 
$Q_{cl}^0(z,t=0)$, to calculate the time evolution of the Husimi distribution, we need to evolve it backward along a trajectory given by,
\begin{align}
    \dot{\zeta}=-i\frac{\partial H^{c*}}{\partial \zeta^*}.
    \label{traj}
\end{align}
$\zeta_0(z,t)$ represents the initial condition of the above trajectory such that $\zeta(t)=z$. Then we need to multiply this by the norm factor, which evolves as,
\begin{align*}
    \dot{w}(z,t)=2\Gamma_I(\zeta_0(z,t))w(z,t),
\end{align*}
with initial condition $w(z,0)=1$. We note that the semiclassical solution of the Husimi distribution coincides exactly with the full quantum Husimi distribution only for Hamiltonians that are quadratic in the creation and annihilation operators $a^\dagger$ and $a$. This condition is not satisfied for either Model~I or Model~II. Consequently, we do not expect exact quantitative agreement between the semiclassical and quantum Husimi dynamics. Nevertheless, our aim is to examine whether the semiclassical description can still capture the essential physics underlying the localization--delocalization transition in these models.
% We start from $|0\rangle$ state whose Husimi representation is given by $Q_{cl}^0(z)=e^{-|z|^2}$. Then we updae it according the the above-mentioned protocol following the trajectory equation \ref {traj}, which in terms of $q,p$ can be written as,
%\begin{align}
 % &\dot{q}  =-\frac{\partial H}{\partial p} + \frac{\partial \Gamma}{\partial q}=(J_R+J_L)\sin p\\
%&\dot{p}  =\frac{\partial H}{\partial q} + \frac{\partial \Gamma}{\partial p}=(J_R-J_L)\cos p - 4\pi\beta V\sin(2\pi \beta q).
%\end{align}
%Where $H$ we denote as the Hermitian part of the Hamiltonian $H_{scl}$. 
\begin{figure*}[t]
    \centering
    \includegraphics[width=0.32\textwidth]{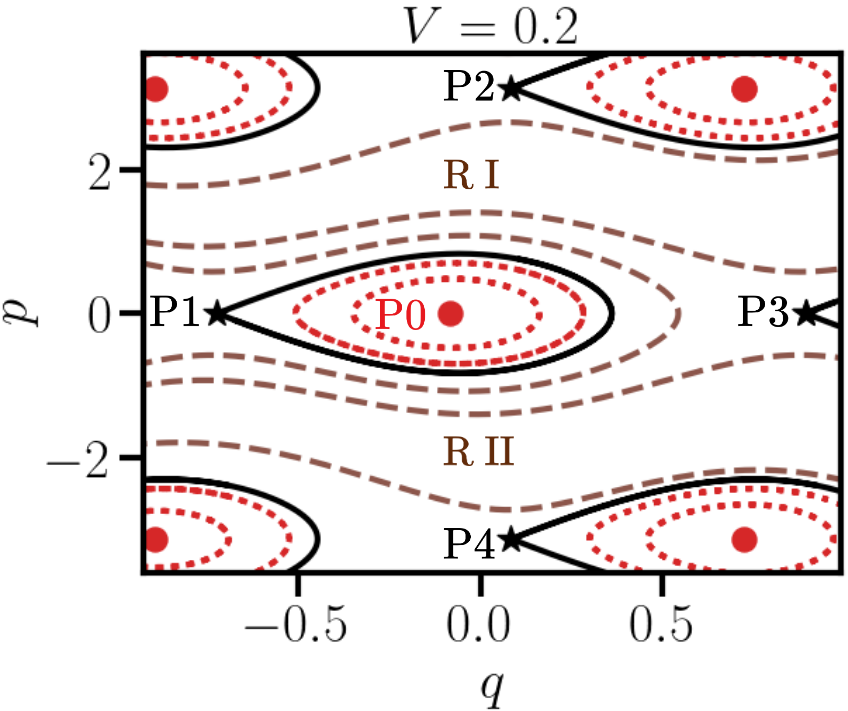}
    \includegraphics[width=0.32\textwidth]{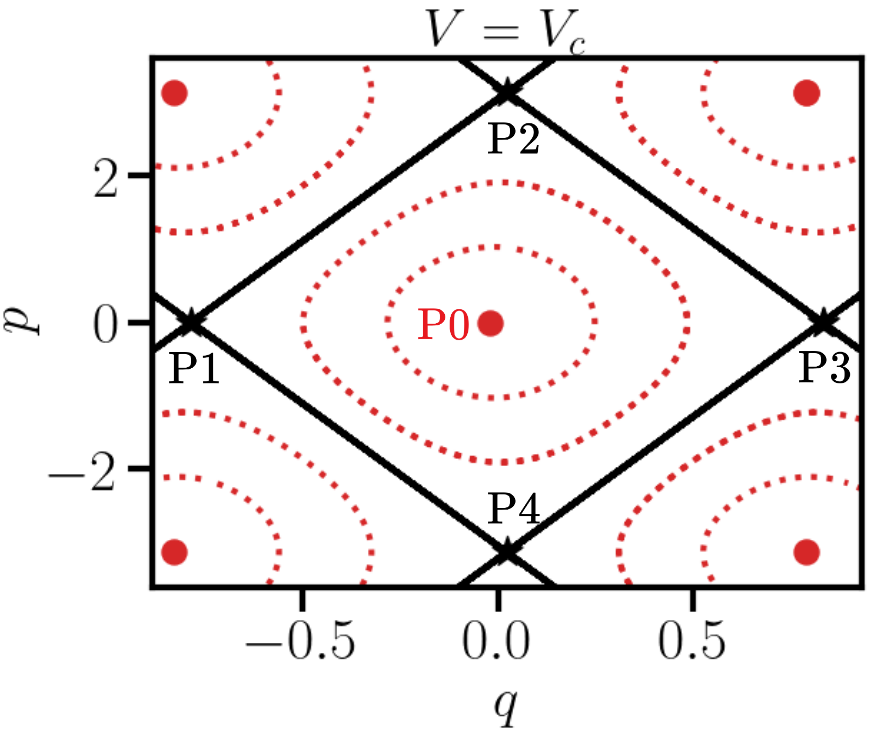}
    \includegraphics[width=0.32\textwidth]{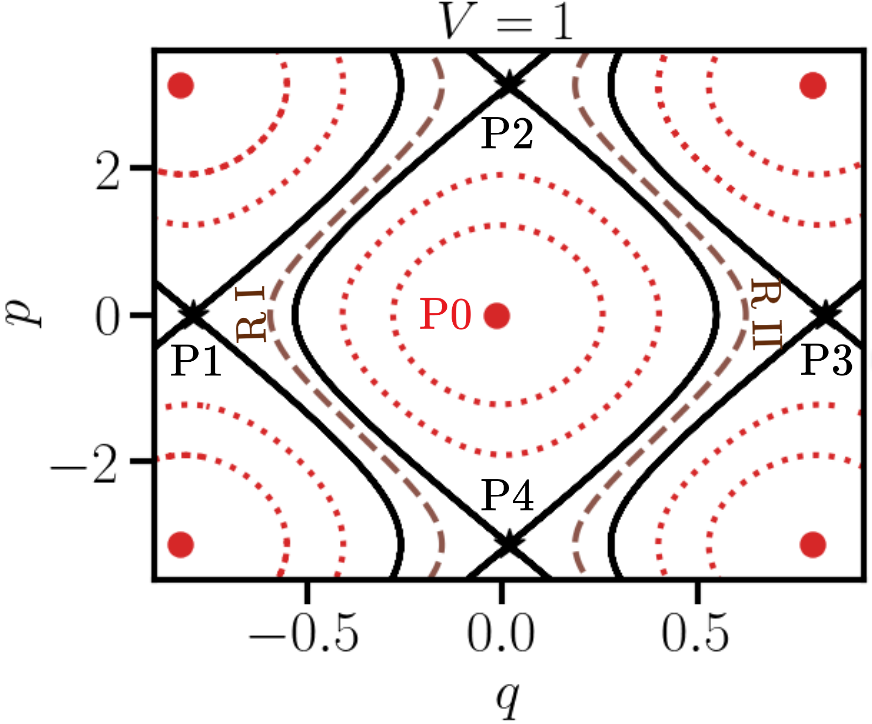}
     \caption{Trajectory evolution goverened by the semiclassical Hamiltonian of Model I [Eq. (\ref{semi_H_model1})] at different $V$ values. Trajectory evolution shows a transition near $V_c=  \frac{1}{2}\sqrt{(J_L+J_R)^2 + \frac{4\Delta^2}{(4\pi\beta)^2}} \sim 0.752$. Data are for $J_L=1, J_R=0.5$. $\beta=\frac{\sqrt{5}-1}{2}$. }
   \label{NH_traj}
  
\end{figure*}
\subsection{Trajectory analysis}
Since the semiclassical time evolution of the Husimi distribution is governed by the underlying phase-space trajectories [Eq.~\eqref{traj}], we perform a detailed analysis of these trajectories using fixed-point analysis. In this context, we like to point out that, in another approach~\cite{Graefe_2010}, canonical equations of motion were derived for a generic non-Hermitian Hamiltonian within a coherent-state approximation. These equations closely correspond to the trajectory equations employed in our semiclassical Husimi dynamics, up to a reversal of the time direction. We have verified that analyzing the trajectories using either formulation leads to the same geometric transition at the level of phase-space dynamics.

\subsubsection{Trajectory analysis of Model I}
First, we replace the operator $\hat{p}$ and $\hat{q}$ with dynamical variables $p$ and $q$ in Eq.~\eqref{continuous_model_I}, 
and identified the classical version of the model I as 
~\cite {Non_hermitian_traj,morice2022quantum},
\begin{equation}
    H_{I}^c=J_R e^{ip} + J_L e^{-ip} + 2V\cos(2\pi\beta q).
    \label{semi_H_model1}
\end{equation}

%In this case, $\Gamma=(J_R- J_L)\sin  p$ is the imaginary part of the Hamiltonian. 
%By $H$ we denote as the Hermitian part of the Hamiltonian $H_{scl}$. 
%We study here the geometry of the trajectories and want to see whether it could mimic the quantum phase transition under the approximate equation of motions given by \cite {Non_hermitian_traj},
%\begin{align}
  %&\dot{q}  =\frac{\partial H}{\partial p} + \frac{\partial \Gamma}{\partial q}=-(J_R+J_L)\sin p\\
%&\dot{p}  =-\frac{\partial H}{\partial q} + \frac{\partial \Gamma}{\partial p}=(J_R-J_L)\cos p + 4\pi\beta V\sin(2\pi \beta q).
%\end{align}
and the trajectory equation Eq.~\eqref {traj} in terms of $q$, $p$ can be written as,
\begin{eqnarray}
   &\dot{q}  =-\frac{\partial \Gamma_R}{\partial p} + \frac{\partial \Gamma_I}{\partial q}=f(q,p)\nonumber \\
&\dot{p}  =\frac{\partial \Gamma_R}{\partial q} + \frac{\partial \Gamma_I}{\partial p}=g(q,p),
\label{dynamical_eq_model1}
\end{eqnarray}
 where, $f(q,p)=(J_R+J_L)\sin p$
 and $g(q,p)=(J_R-J_L)\cos p - 4\pi\beta V\sin(2\pi \beta q)$. 
Here, $\Gamma_R$ refers to the real part of the Hamiltonian $H_{I}^c$,  and $\Gamma_I$ is the imaginary part of the Hamiltonian $H_I^c$. 

Next, our goal is to investigate the phase-space portrait of the dynamical system
described by Eq.~\eqref{dynamical_eq_model1}. Moreover, we examine whether there
exists a critical value $V_c$ such that \(V \ge V_c\) the trajectories remain bounded in
the \(q\) direction, which would imply that the dynamics is localized in position
space. Note that, in the quantum models (both lattice and continuum), there exists such 
a critical value $V = V_c$ that separates the delocalized and localized phases.
Although the dynamical equations can be solved numerically and the phase portrait can be analyzed to identify the critical value $V_c$, it is in general, difficult to pinpoint $V_c$ accurately from numerical simulations alone. We therefore adopt the following strategy: we first determine the fixed points of the dynamical system and analyze the dynamics in their vicinity by linearizing the equations of motion. This allows us to obtain an analytical estimate of $V_c$, which we then compare with the numerical results.

We find the fixed points of the Eq.~\eqref{dynamical_eq_model1}, and they 
are, 
\begin{align*}
    &\dot{q}=0 ~~~\text{gives}~~~ p_0=n\pi, n=0,\pm1,\pm 2\cdots\\
    &\dot{p}=0  ~~~\text{gives}~~~ q_0=\frac{1}{2\pi\beta}\sin^{-1}{\frac{\Delta(-1)^n}{4\pi\beta V}}.
\end{align*}
Note that the fixed point solutions exist only if the following conditions are satisfied,
\begin{align*}
   | \Delta|\le 4\pi\beta V;  ~~~~~\Delta=J_R-J_L.
\end{align*}
We have verified numerically that solutions of Eq.~\eqref{dynamical_eq_model1}
remain unbounded in the \(q\) direction for \(|\Delta| > 4\pi \beta V\).

Linearization of the equations of motion $\dot{q}=f(q,p)$ and $\dot{p}=g(q,p)$ about a fixed point $(q_0,p_0)$ yields the equations governing small fluctuations around the fixed point. Defining $\delta q = q - q_0$ and $\delta p = p - p_0$, the corresponding linearized equations of motion can be written as,\\
\begin{eqnarray}
 \begin{pmatrix} \dot{\delta q} \\ \dot{\delta p}\end{pmatrix}= \begin{pmatrix}\frac{\delta f}{\delta q}& \frac{\delta f}{\delta p} \\
\frac{\delta g}{\delta q} & \frac{\delta g}{\delta p}\end{pmatrix}_{(q_0, p_0)}\begin{pmatrix} {\delta q} \\ {\delta p}\end{pmatrix}=J \begin{pmatrix} {\delta q} \\ {\delta p}\end{pmatrix}. 
\end{eqnarray}

The stability of a fixed point is determined by the nature of the eigenvalues $\lambda$ of the Jacobian matrix $J$. These eigenvalues are obtained from the characteristic equation,
\begin{equation}   \lambda^2=-2(2\pi\beta)^2V(J_L+J_R)\cos{p_0}\cos(2\pi\beta q_0),
\label{Eq: lyponiv model I}
\end{equation}   
Now, for  $p_0=0$ along with $\cos(2\pi\beta q_0)=\sqrt{1-\frac{\Delta^2}{(4\pi\beta V)^2}}$, and $p_0=\pi$ along with $\cos(2\pi\beta q_0)=-\sqrt{1-\frac{\Delta^2}{(4\pi\beta V)^2}}$, the fixed points are stable, as the corresponding eigenvalues are purely imaginary. 
In contrast, for $p_0=0$ along with $\cos(2\pi\beta q_0)=-\sqrt{1-\frac{\Delta^2}{(4\pi\beta V)^2}}$, and $p_0=\pi$ along with $\cos(2\pi\beta q_0)=\sqrt{1-\frac{\Delta^2}{(4\pi\beta V)^2}}$, the fixed points are saddle points, characterized by one positive and one negative real eigenvalue.  
Figure~\ref{NH_traj} clearly demonstrates this behavior. For example, when
\(V = 0.2\), the stable fixed points are indicated by red dots, while the saddle
points are marked by black stars. Trajectories that pass through the saddle
points define the separatrix (shown as the solid black curve), which separates the bounded
from unbounded trajectories.
In the neighborhood of the stable fixed point, the trajectories are closed (identified as red dotted lines in the figures).
In contrast, in regions RI and RII (see Fig.~\ref{NH_traj} ), the orbits are
unbounded in the \(q\) direction, implying delocalization in the spatial
direction. Therefore, in order to identify a critical value \(V_c\) such that
delocalization in the spatial direction is absent, regions RI and RII must shrink
and eventually disappear. This can only occur if there exist separatrix trajectories that
connect the two saddle points.
 Once \(V > V_c\), one expects the separatrix to align in such a way that it
separates bounded closed orbits governed by the stable fixed point from
unbounded orbits in the \(p\) direction, which has been observed for $V=1$ (see Fig.~\ref{NH_traj}  ).  

The essential goal is to determine the critical value \(V = V_c\) such that the
separatrix trajectory passes through the two saddle points (denoted as points
P1 and P2 in Fig.~\ref{NH_traj}). Although we focus on the P1 and P2 points, the
symmetry of the phase portrait makes it straightforward to extend the same
argument to separatrices passing through the P1 and P4 or the P3 and P4 points.

One can combine the dynamical Eq.~\eqref{dynamical_eq_model1} to obtain
\(\mathrm{d}p/\mathrm{d}q = g(q,p)/f(q,p)\). In
general, this equation is too complicated to solve analytically for arbitrary
\(V\). Therefore, we expand the trajectory \(p(q)\) to second order in \(q\) in
the neighborhood of a saddle point, say P1 (which we refer to $(q_0,p_0)$ point in the phase portrait) as,
\[
p(q) = p_0 + m_1 (q - q_0) + \frac{m_2}{2} (q - q_0)^2 .
\]
By matching coefficients order by order in \((q-q_0)\), we obtain explicit expressions
for \(m_1\) and \(m_2\) as a function of $V$,
\begin{align*}
m_1=&\pm \frac{-2(2\pi\beta)^2 \cos(2\pi\beta q_0) V}{(J_R + J_L)\cos p_0}, \\
m_2=&\frac{2}{3m_1}\Bigg[\frac{(2\pi\beta)^3 \sin(2\pi\beta q_0 V)-\frac{1}{2}m_1^2(J_R-J_L)\cos p_0}{(J_R + J_L)\cos p_0}\Bigg] .\nonumber
\end{align*}

It is reasonably straightforward to see that the coefficient \(m_2\)
changes sign at some value \(V = V_*=\frac{1}{2}\sqrt{(J_L+J_R)^2 + \frac{4\Delta^2}{(4\pi\beta)^2}}\) and $m_2$ is zero at this choice of $V$. 

Combining this observation with the behavior of numerically simulated phase
trajectories, we are led to conjecture that \(V = V_*\) plays a special role.
In particular, near this value of \(V\), the curvature of the phase trajectories
changes sign, which intuitively suggests a transition in the nature of the
unbounded regions of the phase portrait—from unbounded motion in the \(q\)
direction to unbounded motion in the \(p\) direction, or vice versa.  This observation motivates us to compute the slope $m_1$ at $V=V_*$. We find that $m_1$ at $V=V_*$ exactly coincides with the slope of the straight line connecting the fixed points P1 and P2 (or equivalently, P1 and P3). Based on this, we first conjecture and then explicitly demonstrate that
\[
p = p_0 + m_1 (q - q_0)
\]
is an exact solution of the differential equation $\mathrm{d}p/\mathrm{d}q = g(q,p)/f(q,p)$ at $V=V_*$. This implies that, for $V=V_*$, the line $p = p_0 + m_1 (q - q_0)$ represents the separatrix trajectory passing through both P1 and P2. Hence, it proves that the critical value of the localization transition $V_c = V_*$.

\begin{figure*}[t]
    \centering
    \includegraphics[width=0.32\textwidth]{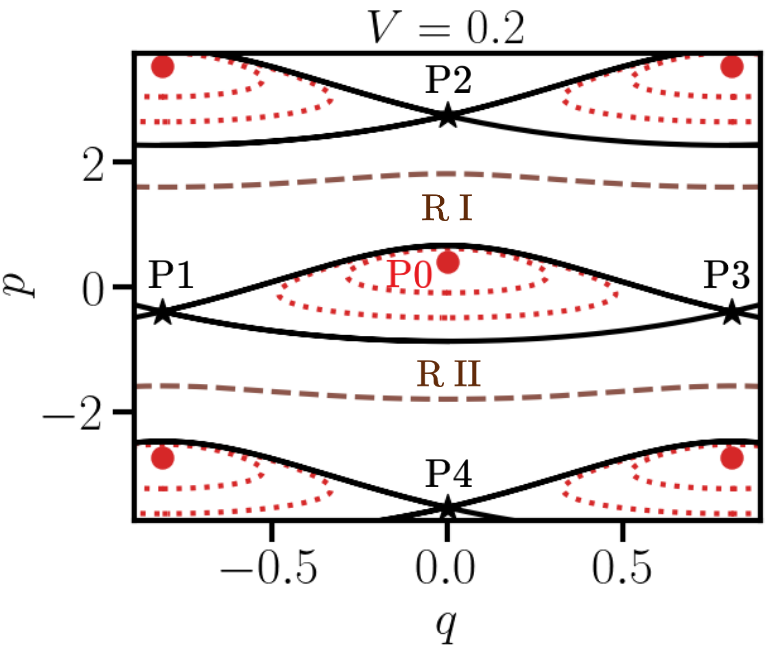}
    \includegraphics[width=0.32\textwidth]{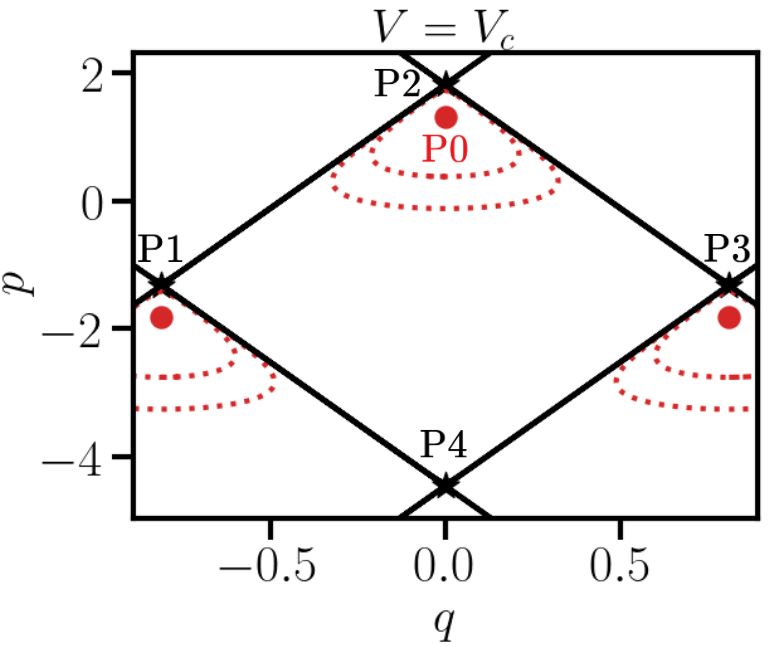}
    \includegraphics[width=0.32\textwidth]{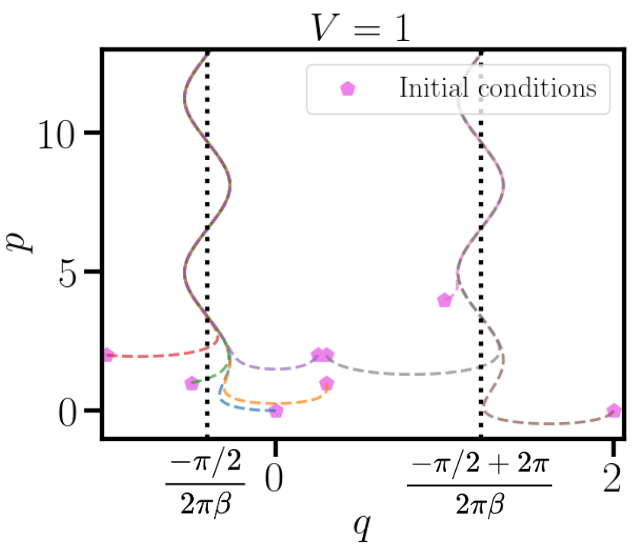}
     \caption{Trajectory evolution governed by the semiclassical Hamiltonian of Model II [Eq. (\ref{semi_H_2})] at different $V$ values. Trajectory evolution shows a transition near $V_c=\frac{2}{\sqrt{1+(2\pi\beta)^2}} \sim 0.498$. Data are for $J=1$. $\beta=\frac{\sqrt{5}-1}{2}$. }
   \label{NH_traj_2}
\end{figure*}

Note that the dynamical equations in Eq.~\eqref{dynamical_eq_model1} are not Hamiltonian equations in the conventional sense. For a usual time-independent Hamiltonian system, each curve in the phase portrait corresponds to a constant energy contour. Consequently, the shrinking of regions RI and RII can be identified by equating the energies of the saddle points, which automatically implies the existence of a separatrix passing through both saddles~\cite{didov2020transport,semiclassical_1}. We employ a similar strategy to determine the transition point for the Hermitian Aubry--André model in the Appendix.~\ref{app_A}.

Applying the same approach to the present case is, however, nontrivial, since the underlying dynamical equations are non-Hamiltonian in nature. Interestingly, we find that the imaginary part of the complex Hamiltonian vanishes at the saddle fixed points. This observation allows us to equate the real parts of the energies at the two saddle points, which correctly reproduces the transition point obtained earlier, i.e., 
\begin{align*}
-(J_L + J_R) + 2V_c \sqrt{1-\frac{\Delta^2}{(4\pi\beta V_c)^2}}
=
(J_L + J_R) \nonumber\\-  2V_c \sqrt{1-\frac{\Delta^2}{(4\pi\beta V_c)^2}},
\end{align*}
which yields
\begin{align*}
V_c=\frac{1}{2}\sqrt{(J_L+J_R)^2 + \frac{4\Delta^2}{(4\pi\beta)^2}}.
\end{align*}

%%%%%%%%%%%%%%%%%%%%%%%%%%%%%%%%%%%%%
We make two interesting observations:
(i) the unbounded orbits in the $q$ direction are also unidirectional (see Region RI and RII in Fig~\ref{NH_traj} for $V=0.2$), which is a signature of the asymmetric hopping terms in the underlying quantum model;
(ii) Unlike the Hermitian Aubry–André model (see Appendix.~\ref{app_A}), the classical transition point in the present case is $\beta$ dependent. We identify a special value of $\beta$ for which the classical and quantum transition points coincide, given by,
\begin{align*}
    \beta=\frac{1}{2\pi\sqrt{7}}; ~~~~\text{with} ~~~~~~J_R=\frac{J_L}{2}, ~~~J_L=1.
\end{align*}
We discuss $\beta=\frac{1}{2\pi\sqrt{7}}$ separately later. \textcolor{black}{We would like to mention here that the special value of $\beta$ for this model depends on the parameter choice $ (J_L, J_R)$ of the model. We find the general expression of the special $\beta$ as,
\begin{align*}
    \beta_s(J_L, J_R)=\frac{1-x}{2\pi \sqrt{4-(1+x)^2}} ; ~~~~~x=\frac{J_R}{J_L};~~~~J_R<J_L.
\end{align*}}
%In the Fig.~\ref{NH_traj}, we present the trajectory evolution data for $\beta=\frac{\sqrt{5}-1}{2}$. For large $V=1$, we see closed orbits because of the initial condition slightly far from the stable fixed points, as well as extended orbits in $p$ direction because of the initial conditions slightly far from the saddle fixed points, these two different kind of trajectories are separated by the seperatrices marked by the black solid line, at $V=V_c$, the seperatrices merge together, on the other hand for small $V=0.2$ we see closed orbits because of the initial condition slightly far from the stable fixed points, as well as extended orbits in $q$ direction because of the initial conditions slightly far from the saddle fixed points  Which represents a geometrical delocalization-localization transition at $V=V_c$.

%We mention here that the trajectory evolutions are quite richer in the $J_L \ne J_R$ case as compare to the Hermitian $J_L=J_R$ case(see Appendix.~\ref{app_A} Fig.~\ref{H_traj}), the effect of non-hermiticity is there in the trajectories also as it predicts the unidirectional nature of the motion for $V<V_c$, but it fails to predict the quantum transition point.

\subsubsection{Trajectory analysis of Model II}
The semiclassical version of Model II can be obtained by replacing the operators $\hat{q}$ and $\hat{p}$ in Eq.~\eqref{continuous_model_II} with the corresponding classical variables $q$ and $p$, respectively.
   \begin{align}
       H_{II}^c=2\cos{p} + V e^{-2\pi i \beta q}.
       \label{semi_H_2}
   \end{align}
   We take $J=1$.
   The trajectory equation  (Eq.~\eqref{traj}) reads as,
\begin{align}
  &\dot{q}  =2\sin p - 2\pi\beta V\cos(2\pi \beta q) \nonumber \\
&\dot{p}  = - 2\pi\beta V\sin(2\pi \beta q) \nonumber .
\end{align}
We follow the same strategy as before and identify the fixed points for $\pi \beta V \leq 1$ as,
\begin{align*}
    &\dot{p}=0 ~~~\text{gives}~~~ q_0=n\pi/(2\pi\beta), ~~~~~~n=0,\pm1,\pm 2\cdots,\\
    &\dot{q}=0  ~~~\text{gives}~~~ p_0=\sin^{-1}{[\pi\beta V(-1)^n]}.
\end{align*}
The eigenvalues of the Jacobian matrix $J$ are given by the following equation,
\begin{align*}
   \lambda^2=-2(2\pi\beta)^2V\cos{p_0}\cos(2\pi\beta q_0).
\end{align*}
For the combinations $q_0=0, \cos( p_0)=\sqrt{1-(\pi\beta V)^2}$ and $q_0=\pi/(2\pi\beta), \cos( p_0)=-\sqrt{1-(\pi\beta V)^2}$  we have stable fixed points as the roots are completely imaginary, on the other hand for, $q_0=0, \cos( p_0)=-\sqrt{1-(\pi\beta V)^2}$ and $q_0=\pi/(2\pi\beta), \cos( p_0)=\sqrt{1-(\pi\beta V)^2}$  we have saddle fixed points as we have one positive real and another negative real eigenvalue.
\begin{figure*}
    \centering
    \includegraphics[width=0.85\textwidth, height=0.65\textwidth]{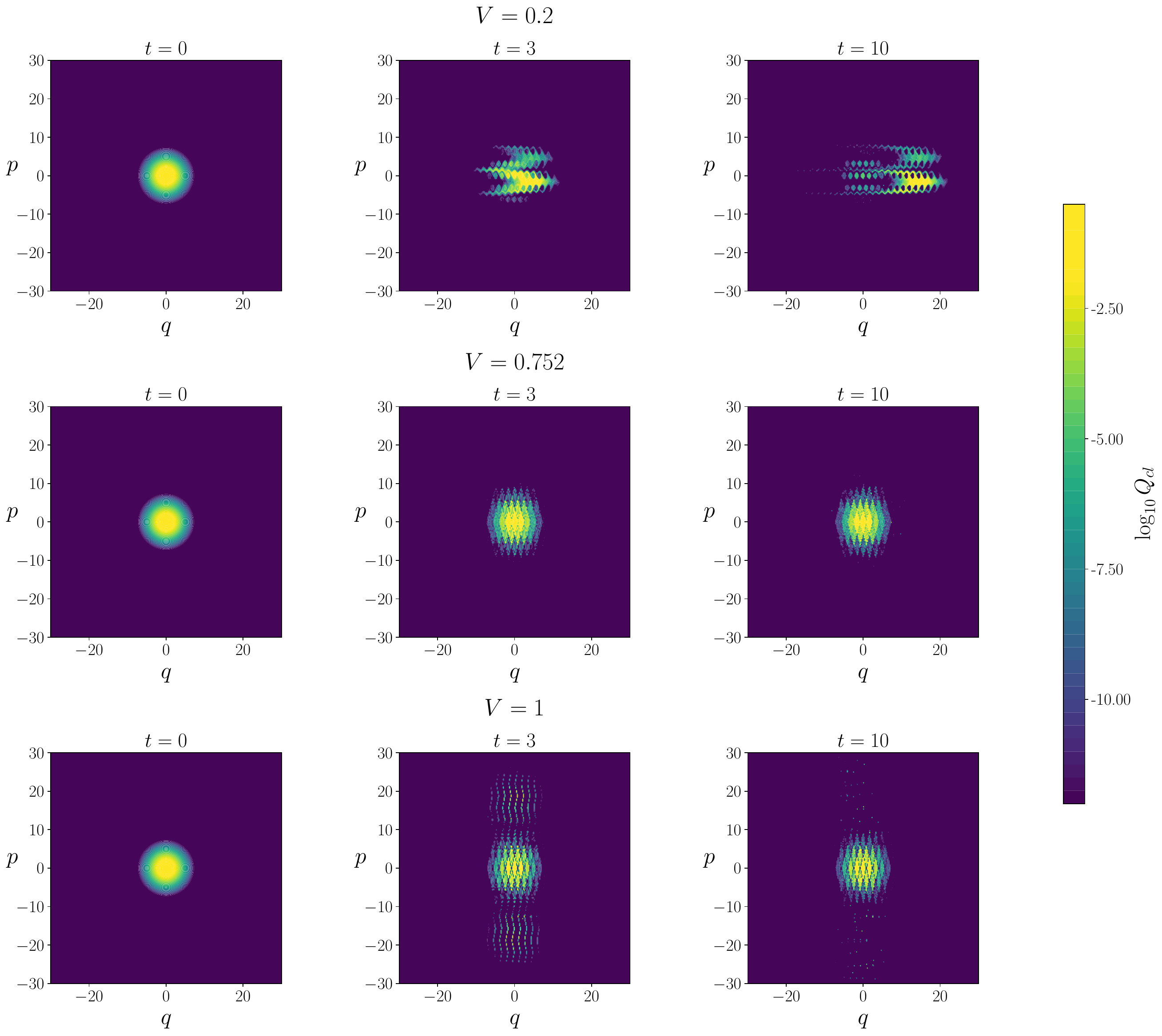}
    
    \caption{Semiclassical Husimi distribution at different time instances starting from coherent state $|z=0\rangle$ and evolving under the Hamiltonian [Eq. (\ref{semi_H_model1})]. Data is for $V=0.2$, $0.752, 1$. The color code represents the Husimi values. }
   \label{C1}
\end{figure*}
Next, we again perform a rigorous analysis, following the same procedure as before, and analytically determine the transition point as
\begin{align*}
    V_c=\frac{2}{\sqrt{1+(2\pi\beta)^2}},
\end{align*}
which is also $\beta$ dependent. 
In Fig.~\ref{NH_traj_2}, we present the trajectory evolution for $\beta = \frac{\sqrt{5}-1}{2}$. The geometrical changes visible in the trajectories signal a transition, which agrees well with the analytically predicted transition point obtained from the saddle-point analysis.
Once again, we find a special value of $\beta = \frac{\sqrt{3}}{2\pi}$ for which the transition point is $V_c = 1$, in agreement with the corresponding quantum transition point. 

%\textcolor{red}{The large $V>>V_c$ behaviour could also be understand with simple analytics. In the potential dominating limit, the equation of motion reads as,
%\begin{align*}
  %  \dot{q}\simeq-2\pi\beta V\cos(2\pi\beta q),\\
  %  \dot{p}\simeq-2\pi\beta V\sin(2\pi\beta q).
%\end{align*}
%As we see $\dot{q}$ is a function of $q$ only, this could be easily solved, and one can get,
%\begin{align*}
 %   |\sec(2\pi\beta q(t)) + \tan(2\pi\beta q(t))|=Ae^{-(2\pi\beta)^2 Vt},
%\end{align*}
%where $A$ is the integration constant that depends on the initial condition ($q_0, p_0$).
%At a large enough $V$ and $t$ we can write,
%\begin{align*}
 %   &|\sec(2\pi\beta q(t)) + \tan(2\pi\beta q(t))|\simeq0,\\
  %  &1+\sin(2\pi\beta q(t))=0,\\
   % &q(t)=\frac{-\pi/2 + 2n\pi}{2\pi\beta},~~~~ \text{where} ~~~~n=0,\pm1, \pm 2,\cdots.
%\end{align*}
%Hence, for large enough $V$, and long-time $q(t)$ pins to the above-mentioned quantized values, whichever initial conditions we start with. And for the given $q(t)=\frac{-\pi/2 + 2n\pi}{2\pi\beta}$, one easily get,
%\begin{align*}
 %   p(t)=p_0 + 2\pi\beta V t.
%\end{align*}
%In Fig.~\ref{NH_traj_2}, right-most panel, we can see this behaviour even for $V=1$. In large time $t$, $q(t)$ gets quantized, and we see unbounded motion in the $p$ direction oscillating around the quantized $q(t)$. We mark the quantized values of $q(t)$ by black dotted lines. 
%}
The large-$V$ ($V \gg V_c$) behavior can also be understood analytically. 
In the potential-dominated limit  $V\to \infty$, the equations of motion reduce to
\begin{align*}
    \dot{q} &\simeq -2\pi\beta V \cos(2\pi\beta q), \\
    \dot{p} & = -2\pi\beta V \sin(2\pi\beta q).
\end{align*}
Since $\dot{q}$ depends only on $q$, the first equation can be solved straightforwardly, yielding
\begin{align*}
    \left|\sec(2\pi\beta q(t)) + \tan(2\pi\beta q(t))\right|
    = A e^{-(2\pi\beta)^2 V t},
\end{align*}
where $A$ is an integration constant determined by the initial condition $(q_0,p_0)$.
For sufficiently large $V$ and long times $t$, the right-hand side vanishes exponentially, and we obtain
\begin{align*}
    &\left|\sec(2\pi\beta q(t)) + \tan(2\pi\beta q(t))\right| \simeq 0,\\
    & 1 + \sin(2\pi\beta q(t)) = 0.
    \end{align*}

so that
\begin{align*}
    q(t) = \frac{-\pi/2 + 2n\pi}{2\pi\beta}, 
    \qquad n = 0, \pm 1, \pm 2, \ldots.
\end{align*}
Thus, for sufficiently large $V$ and long times, $q(t)$ becomes pinned to these discrete values, irrespective of the initial condition.
Substituting $q(t) = \frac{-\pi/2 + 2n\pi}{2\pi\beta}$ into the equation for $\dot{p}$, we obtain
\begin{align*}
    p(t) = p_0 + 2\pi\beta V t,
\end{align*}
indicating unbounded linear growth in the $p$ direction.
In the rightmost panel of Fig.~\ref{NH_traj_2}, this behavior is already visible even for $V=1$. At long times, $q(t)$ approaches the quantized values given above, while $p(t)$ exhibits unbounded motion, oscillating around these pinned values of $q(t)$. The quantized values of $q(t)$ are indicated by black dotted lines.

%In the Fig.~\ref{NH_traj_2}, we present the trajectory evolution data for $\beta=\frac{\sqrt{5}-1}{2}$, and the geometrical changes, which are visible from the figure, represent a transition, and that matches well with the analytically predicted transition point from the saddle point analysis. For small $V=0.2$, we see closed orbits because of the initial condition slightly far from the stable fixed points, as well as extended orbits in $q$ direction because of the initial conditions slightly far from the saddle fixed points, these two different kind of trajectories are separated by the seperatrices marked by the black solid line, at $V=V_c$, the seperatrices merge together, on the other hand for large $V=1$ as predicted from the analytics there are no fixed points, we see extended orbits in the $p$ direction, which represents a geometrical delocalization-localization transition at $V=V_c$.
\begin{figure}[!h]
    \centering
    \includegraphics[width=0.45\textwidth]{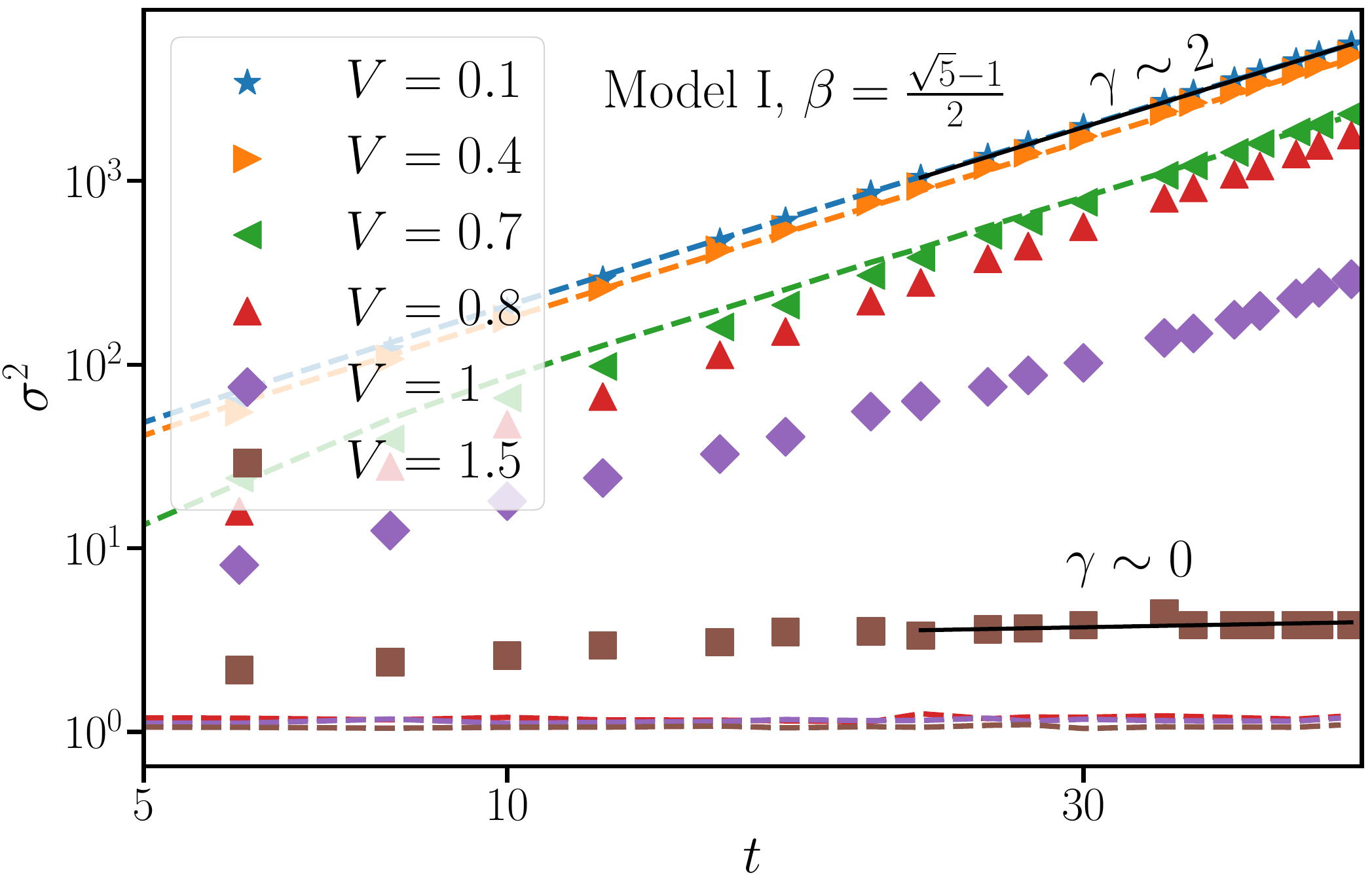}
     \includegraphics[width=0.45\textwidth]{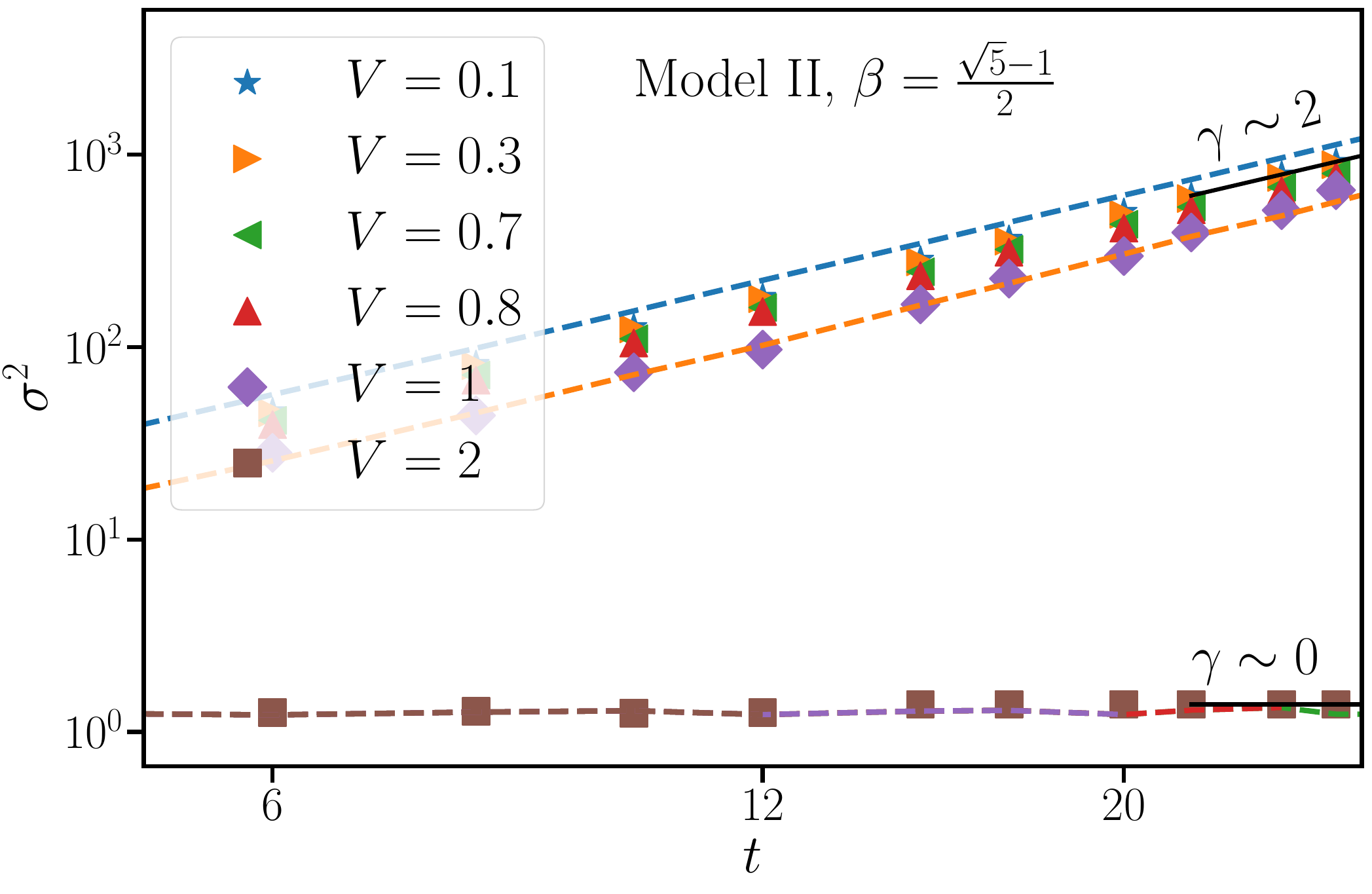}
      \caption{$\sigma^2$ vs. $t$ plots for the quantum continuous and semiclassical Husimi dynamics for Model I(upper panel) and Model II(lower panel). Points represent the quantum continuous data, while dashed lines represent the data extracted from the classical Husimi spreading.  The black solid line represents fitting to a function $\sigma^2 \sim t^{\gamma}$.}
\label{class_qunaum_comp}
\end{figure}
\subsection{Time evolution of the semiclassical Husimi distribution}
Next, we study the time evolution of the semiclassical Husimi distribution governed by Eq.~\eqref{semi_classical_soln} for different values of $V$ at various time instances, for both Model~I and Model~II. We choose an initial coherent state $|z_0\rangle = |z=0\rangle$, such that the initial Husimi distribution is given by
\begin{equation}
Q_{\mathrm{cl}}^{0}(z,t=0) = e^{-|z|^{2}} \, .
\end{equation}

\begin{figure}[!h]
    \centering
    \includegraphics[width=0.45\textwidth]{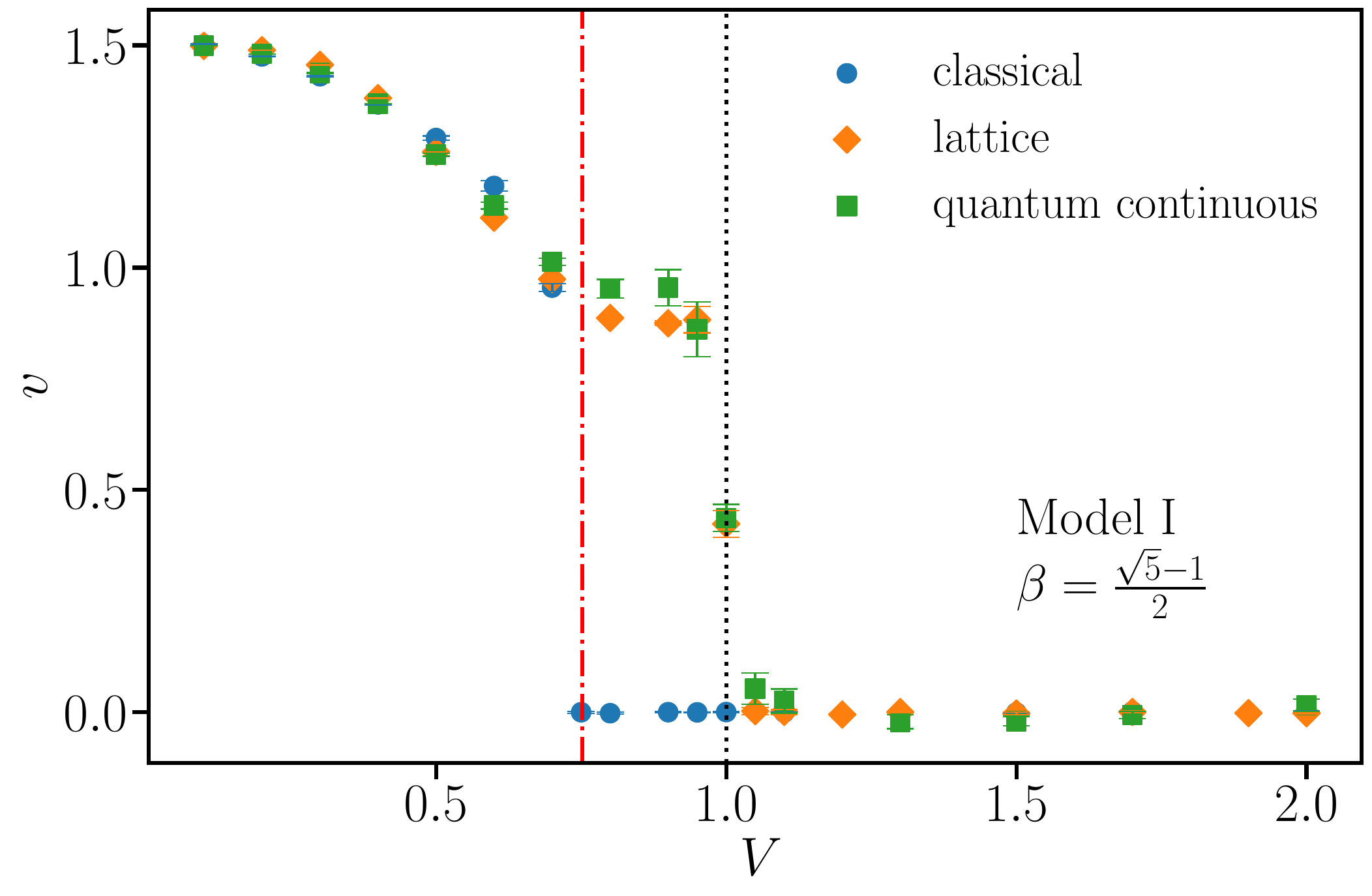}
         \includegraphics[width=0.45\textwidth]{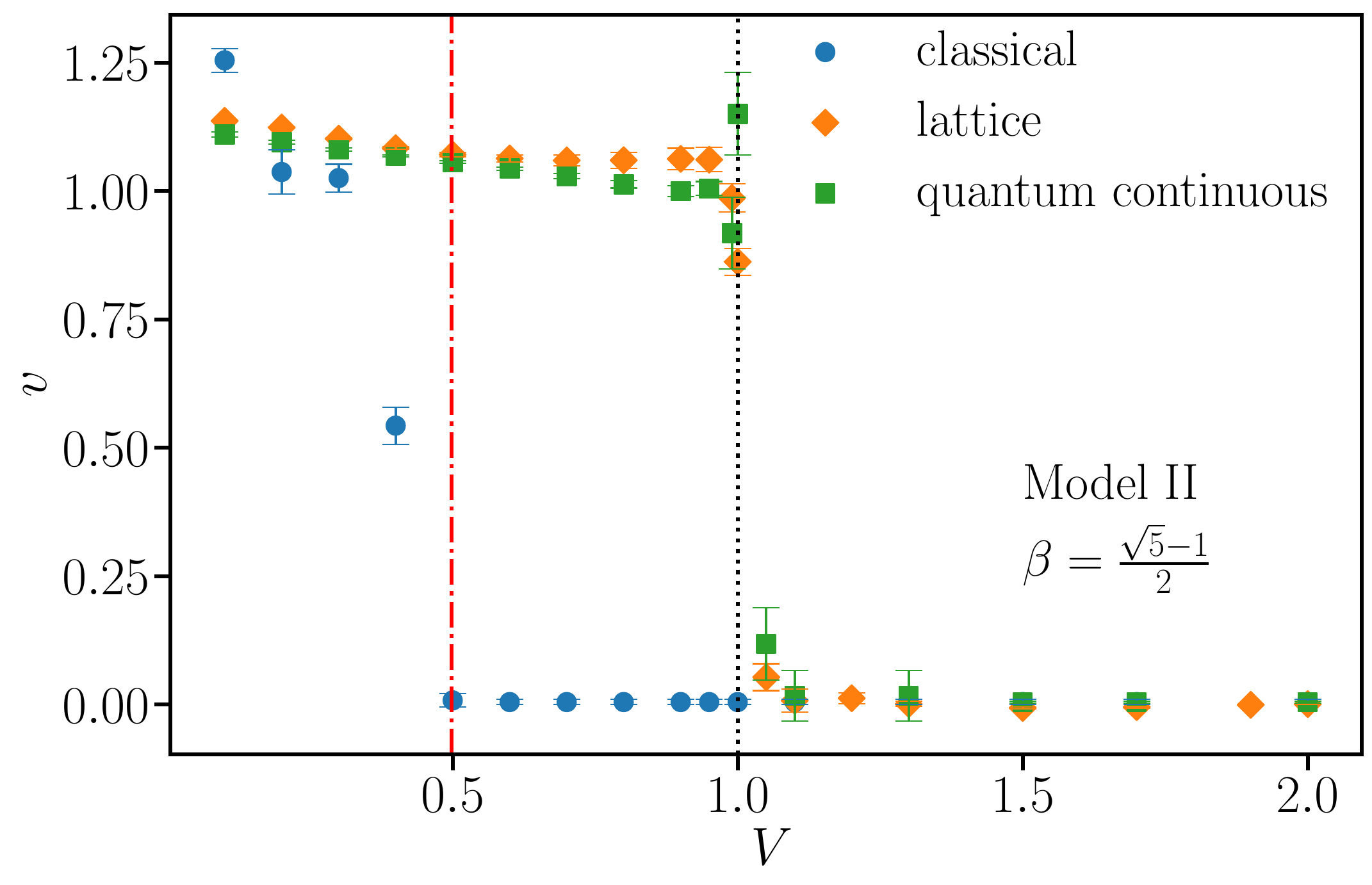}

    \caption{Upper panel: Comparison of the velocity of the excitation of the lattice Model I~{Eq. (\ref{Model_1_lattice})] with the excitation velocity extracted from the Husimi distribution of the quantum continuous Model I~[Eq. (\ref{continuous_model_I})], as well as the velocity extracted from the semiclassical Husimi evolution.  Lower panel: Comparison of the velocity of the excitation of the lattice Model II~[Eq. (\ref{Model_2_lattice})] with the excitation velocity extracted from the Husimi distribution of the quantum continuous Model II~[Eq. (\ref{continuous_model_II})] and the velocity extracted from the corresponding semiclassical Husimi dynamics. We take $\beta=(\sqrt{5}-1)/2.$   \textcolor{black}{Results shown here are extrapolated thermodynamic findings}. }
    \label{comp_actual_beta_model_1}
}
\end{figure}

In Fig.~\ref{C1}, we show the time evolution of the semiclassical Husimi distribution for Model~I at $V=0.2$, $0.752$, and $1$, respectively. As expected, the Husimi dynamics follows the underlying classical trajectories, modulated by the corresponding norm factor. From our earlier analysis of the classical trajectory dynamics, we know that a  transition occurs at $V_c \simeq 0.752$ for $\beta = (\sqrt{5}-1)/2$, $J_L = 1$, and $J_R = 0.5$. Consistently, the semiclassical Husimi dynamics also signals a transition near $V \simeq 0.752$, as evidenced by the cessation of spreading of the distribution along the $q$-direction. We have verified that this conclusion remains unchanged even when the initial state is not chosen to be a coherent state.

\textcolor{black}{
Next, we compare the second moment obtained from the classical and quantum Husimi distributions in Fig.~\ref{class_qunaum_comp} (upper panel). The black solid line denotes a fit of the form $\sigma^2 \sim t^\gamma$. For the quantum continuum model, we have seen that $\gamma \simeq 2$ for $V \leq V_c = J_L$. In contrast, for the semiclassical dynamics, we find $\gamma \simeq 2$ only for $V < 0.752$, while for $V \geq 0.752$, $\gamma \simeq 0$, indicating localization.
A similar analysis was performed for Model II. While the corresponding Husimi distributions are presented in Appendix~\ref{Model_II_Husimi}, the scaling of $\sigma^2$ is shown in the lower panel of Fig.~\ref{class_qunaum_comp}, and we see $\gamma$ drops to zero around $V \sim 0.498$. For both Model I and Model II, the semiclassical and quantum continuum results are in reasonable agreement in the delocalized regime at small $V$. However, the critical potential strength predicted by the semiclassical dynamics differs from that of the quantum model. Since, for our choice of parameters, the semiclassical critical value $V_c$ is smaller than the quantum prediction, the agreement deteriorates once $V$ exceeds the semiclassical $V_c$. 
}

Further, we compare the dynamical properties of the lattice model—quantified previously via the speed of propagation—with those obtained from the time evolution of the Husimi distribution of the corresponding continuous quantum model, as well as from the semiclassical Husimi dynamics. As discussed earlier and shown in the upper panel of Fig.~\ref{comp_actual_beta_model_1}, for $\beta = (\sqrt{5}-1)/2$ the continuous model exhibits dynamical behavior very similar to that of its lattice counterpart, with both showing a transition near $V \simeq 1$. In contrast, the speed of propagation extracted from the semiclassical Husimi distribution drops already near $V_c \sim 0.752$, coinciding with the merging of separatrices in the classical phase space. \textcolor{black}{Red dashed line represents the semiclassical delocalization-localization transition point, while black dotted line represents the transition point for the lattice as well as its continuous version. The results presented are extrapolated thermodynamic data as discussed in the Appendix~\ref {extrapolated_results}. Similarly, for the second model, lattice and continuous quantum models exhibit a transition near $V \simeq 1$, the semiclassical description fails to reproduce this behavior (see Fig.~\ref{comp_actual_beta_model_1} lower panel). Instead, the speed of propagation obtained from the semiclassical Husimi evolution drops already near $V \sim 0.498$, coinciding with a transition in the classical trajectories. As in Model~I, the classical phase transition point systematically underestimates the quantum localization transition.} In summary, both the semiclassical Husimi dynamics and the classical trajectory analysis systematically underestimate the quantum localization transition point. \textcolor{black}{ Unlike the lattice model and its continuous version, the semiclassical propagation yields zero speed of propagation at the transition point.
}

\begin{figure}
    \centering
    \includegraphics[width=0.45\textwidth]{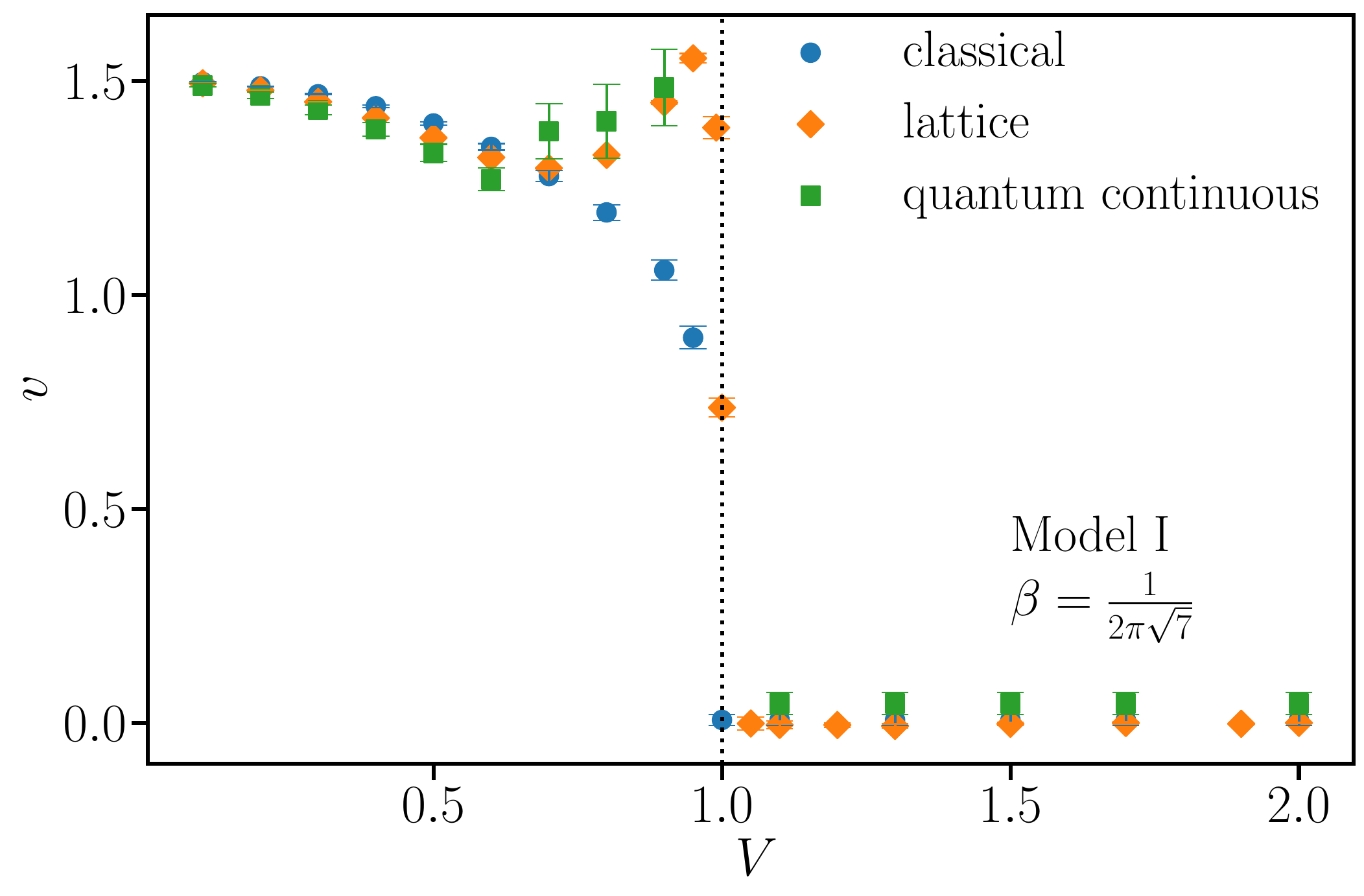}
    \includegraphics[width=0.45\textwidth]{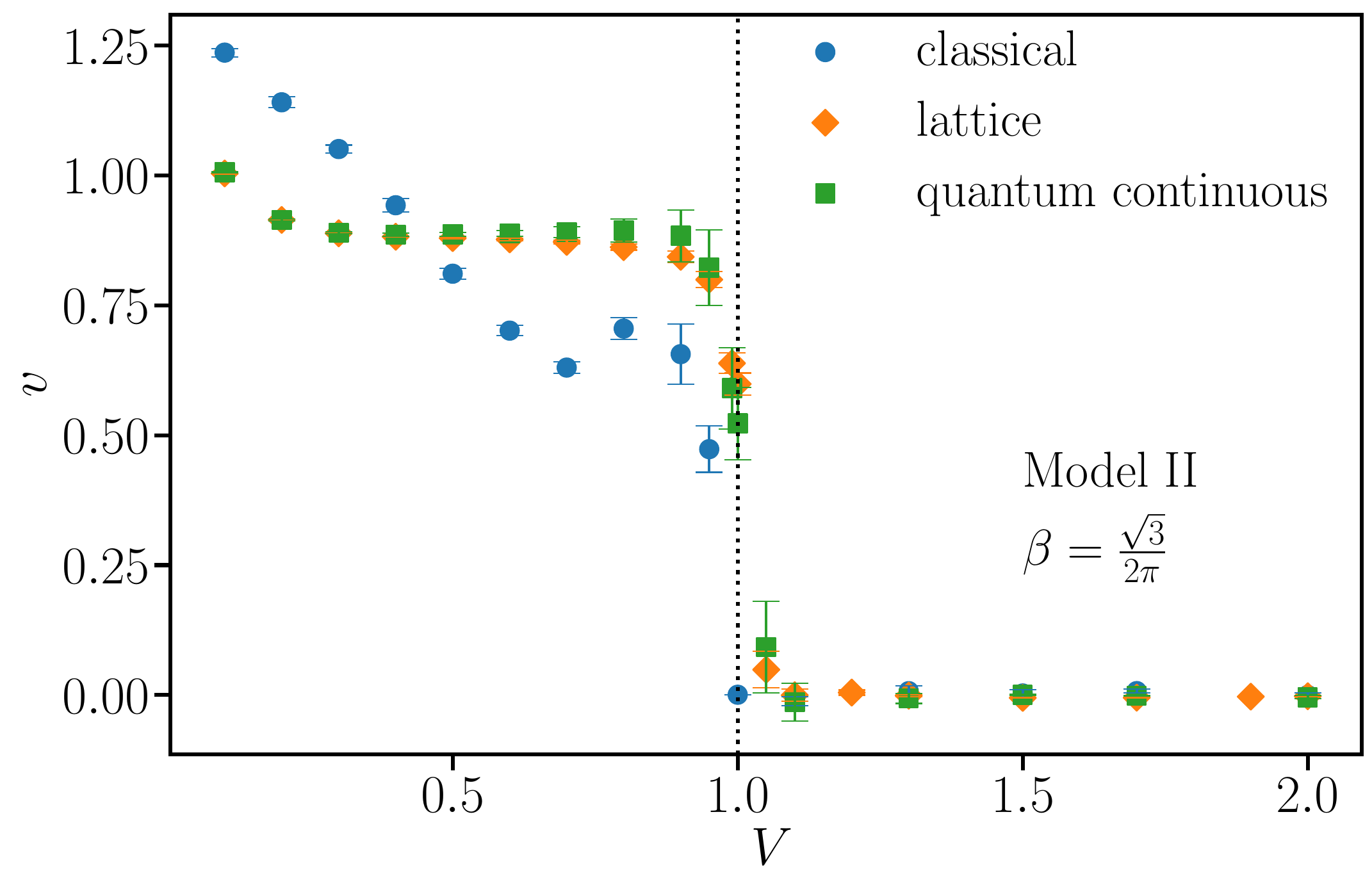}
    \caption{Upper panel: Comparison of the velocity of the excitation of the lattice Model I [Eq. (\ref{Model_1_lattice})] with the excitation velocity extracted from the Husimi distribution of the quantum continuous Model I~[Eq. (\ref{continuous_model_I})],as well as the velocity extracted from the semiclassical Husimi evolution for $\beta=\frac{1}{2\pi\sqrt{7}}$. Lower panel: Comparison of the velocity of the excitation of the lattice Model II~[Eq. (\ref{Model_2_lattice})] with the excitation velocity extracted from the Husimi distribution of the quantum continuous Model II~[Eq. (\ref{continuous_model_II})] and the velocity extracted from the corresponding semiclassical Husimi dynamics for $\beta=\sqrt{3}/(2 \pi)$. \textcolor{black}{Results shown here are extrapolated thermodynamic findings}. } 
    \label{comp_modified_beta}
\end{figure}

\paragraph*{Special $\beta$ limit:}
  For a special choice of the irrational parameter, namely $\beta=\frac{1}{2\pi\sqrt{7}}$ for Model~I and $\beta = \frac{\sqrt{3}}{2\pi}$ for Model~II (for our chosen set of parameters), we find that the dynamical and the corresponding geometrical phase transition points are qualitatively the same. \textcolor{black}{The extrapolated thermodynamic data are compared in Fig.~\ref{comp_modified_beta}. As can be seen, for both models, the classical and quantum cases (including both the continuum and lattice realizations) exhibit a transition at $V=1$. Beyond this point ($V>1$), the propagation velocity obtained from the appropriate extrapolation procedure (see Appendix~\ref{extrapolated_results}) approaches zero, i.e., $v \simeq 0$.
However, an important difference remains at the critical point itself. While the propagation velocity vanishes exactly at $V=1$ in the classical models ($v=0$), it remains finite in the corresponding quantum models. This distinction persists even in the thermodynamic limit. In this regard, the Hermitian AA model behaves differently. At the transition point, $V=1$, the propagation velocity vanishes, $v=0$, for both the classical and quantum models. Therefore, the distinction observed between the classical and quantum non-Hermitian models at criticality is absent in the Hermitian AA model (see Appendix~\ref{app_A} for details).
We would also like to clarify the situation for the quantum continuum model. Simulations at longer times are computationally challenging due to the exponential growth of the accessible harmonic-oscillator Hilbert space. Consequently, the available finite-time data at $V=1$ are insufficient to obtain a reliable extrapolation for Model I. For this reason, we do not report the extrapolated value at $V=1$ in the upper panel of Fig.~\ref{comp_modified_beta}.
}

\begin{figure*}
    \centering
    \includegraphics[width=1\textwidth]{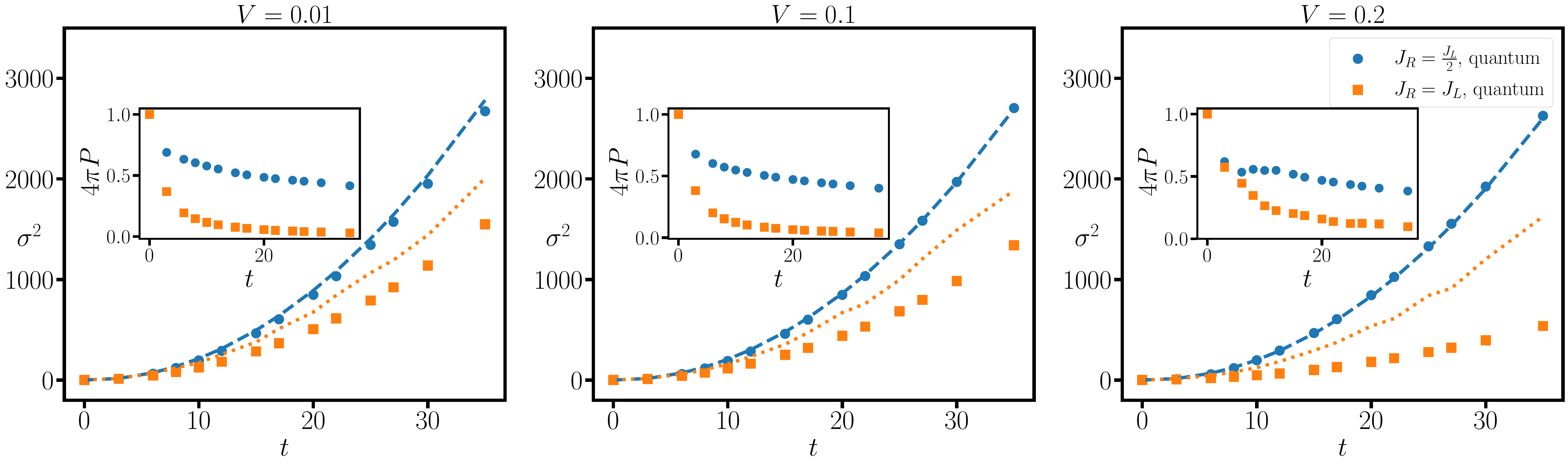}
   \caption{Main panel: Comparison of $\sigma^2$ extracted from Quantum and classical Husimi distribution in the limit $\beta V \to \frac{\Delta}{4\pi}$, both for non-Hermitian(Model I with $J_R=\frac{J_L}{2}$) and Hermitian(Model I with $J_R=J_L$). The inset shows the change in the purity with time of the quantum Husimi distribution in the above mentioned limit. Square and circle points represent the data extracted from quantum Husimi with $J_L=J_R$ and $J_R=\frac{J_L}{2}$, respectively, and dotted lines represent the classical data for $J_L=J_R$, dashed lines represent the classical data with$J_R=\frac{J_L}{2}$.}
   \label{Purity_1}
\end{figure*}

\paragraph*{Phase space purity:} Moreover, to quantify how effectively the classical description mimics the quantum dynamics, we compute the phase-space purity, defined as
\begin{equation}
    P(t) = \int \left| Q_q^{\mathrm{norm}}(q,p,t) \right|^2 \, dp \, dq,
    \label{Eq:purity}
\end{equation}
where
\[
Q_q^{\mathrm{norm}}(q,p,t)
= \frac{Q_q(q,p,t)}
{\int Q_q(q,p,t) \, dp \, dq}.
\]

At $t=0$, we initialize the system in a coherent state,
\[
Q_q(q,p,0) = e^{-(q^2 + p^2)/2},
\]
which yields
\[
P(0) = \frac{1}{4\pi}.
\]
As time evolves, if $P(t)$ does not decay rapidly, this indicates that the classical description continues to approximate the quantum dynamics reasonably well within that parameter regime.

 %See Figs.~\ref{Purity_1} and \ref{Purity_2}.

%\begin{figure*}
 %   \centering
  %  \includegraphics[width=1\textwidth]{Co_2.pdf}
   %\caption{Purity for Model II}
    %\label{Purity_2}
%\end{figure*}

In the context of quantum--classical correspondence, the relevant timescale is typically the Ehrenfest time, 
\begin{equation}
t_E \sim \lambda^{-1},
\end{equation}
which determines the duration over which quantum dynamics can be reasonably approximated by classical evolution, with $\lambda$ denoting the effective Lyapunov exponent (see Eq.~\eqref{Eq: lyponiv model I}). Up to this time, semiclassical descriptions remain valid, whereas beyond it genuine quantum interference effects dominate the dynamics.

For the Hermitian case, it is clear from Eq.~\eqref{Eq: lyponiv model I} (when $J_R=J_L$) that for a given potential strength $V$, the regime of large Ehrenfest time corresponds to the limit $\beta \to 0$. In this limit, $t_E$ becomes extremely large, and  presumably the classical--quantum correspondence can be formally well established. However, this limit is of limited practical relevance. To observe significant quasiperiodic effects in a finite-size lattice, one requires $\beta$ to be reasonably large. Increasing $\beta$, however, reduces the Ehrenfest time and consequently shrinks the temporal window over which classical dynamics faithfully reproduces the quantum evolution. Thus, in the Hermitian case, the classical limit and the physically relevant quasiperiodic regime do not strongly overlap, restricting the practical utility of the semiclassical approximation in dynamics. 

The situation changes qualitatively in the non-Hermitian case. There,  for model I the regime of large Ehrenfest time corresponds to the condition
\begin{equation}
\beta V \to \frac{\Delta}{4\pi}, 
\end{equation}
for which $\lambda \to 0$ (see Eq.~\eqref{Eq: lyponiv model I}).
For fixed and sufficiently small $V$, this allows $\beta$ to take reasonably large values while the system remains in the delocalized phase. Hence, unlike the Hermitian case, the semiclassical regime overlaps with a physically meaningful parameter window.

We demonstrate this explicitly in Fig.~\ref{Purity_1}. In this limit, the purity for non-Hermitian system, does not decay rapidly, indicating that quantum interference effects are significantly suppressed. Moreover, the wavepacket width $\sigma(t)$ can be reproduced with high accuracy using classical evolution based on the Husimi distribution. As shown in the figure, the quantum results (dots) agree remarkably well with the classical Husimi calculations (dashed lines) up to a substantial time scale.

Figure~\ref{Purity_1} further highlights a crucial contrast: in the same $\beta$ regime, the Hermitian model fails to reproduce the quantum dynamics even at extremely short times. The classical approximation breaks down almost immediately, implying that the effective Ehrenfest time is very small. This stark difference underscores the qualitative impact of non-Hermiticity.

\begin{figure*}[t]
    \centering
    \includegraphics[width=0.32\textwidth]{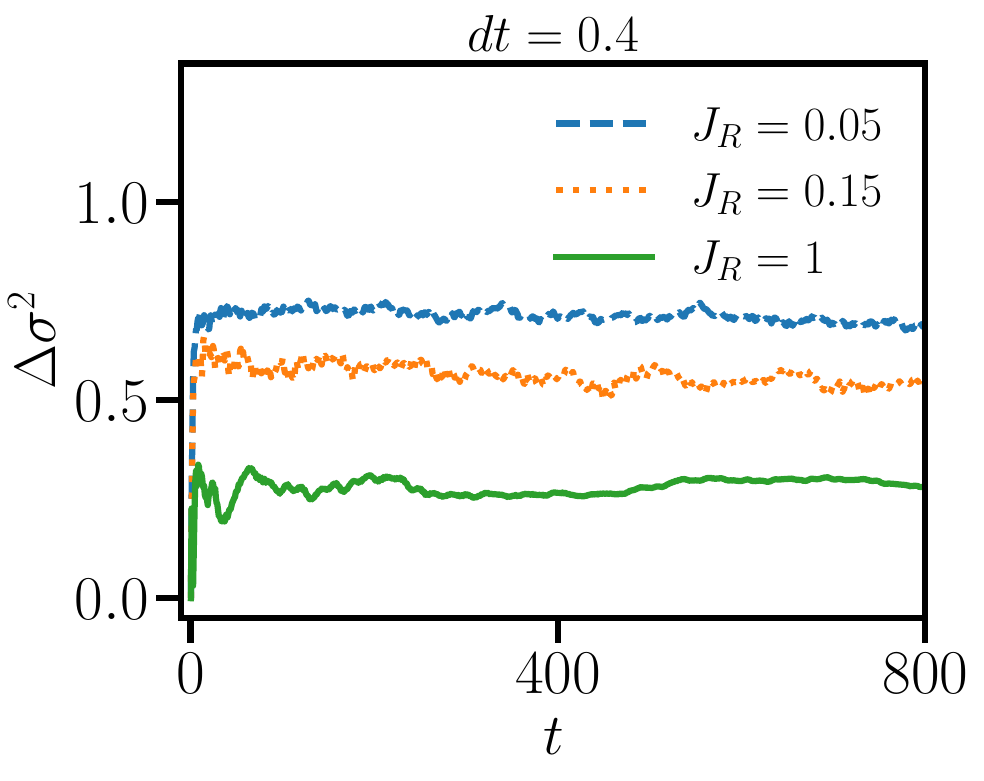}
    \includegraphics[width=0.32\textwidth]{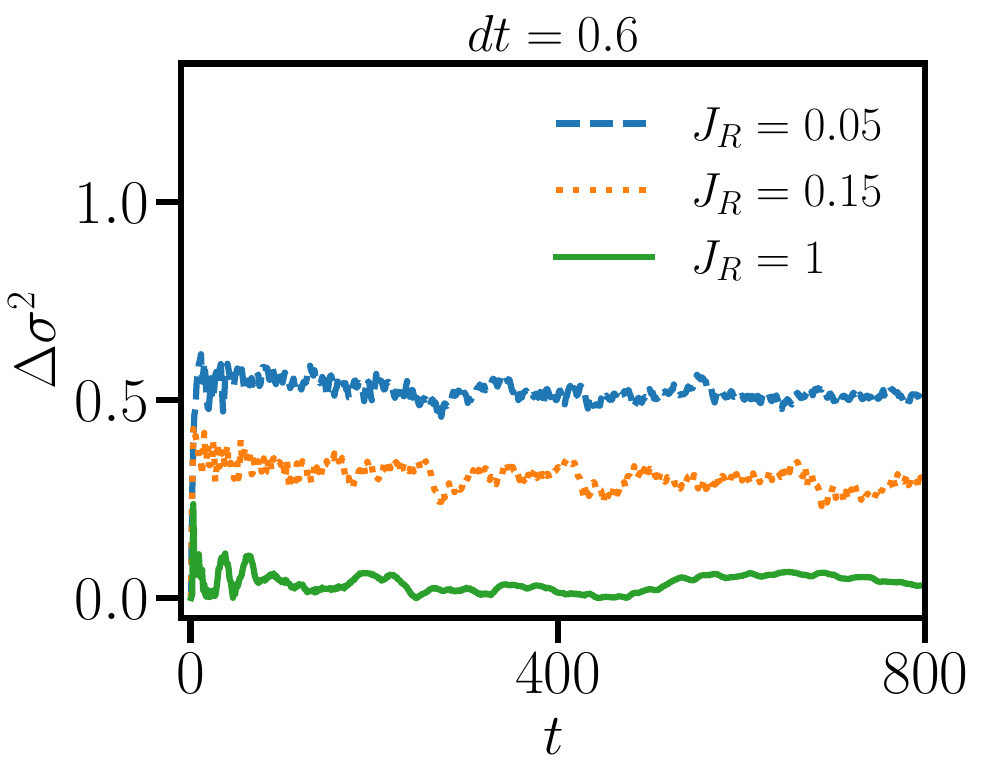}
    \includegraphics[width=0.32\textwidth]{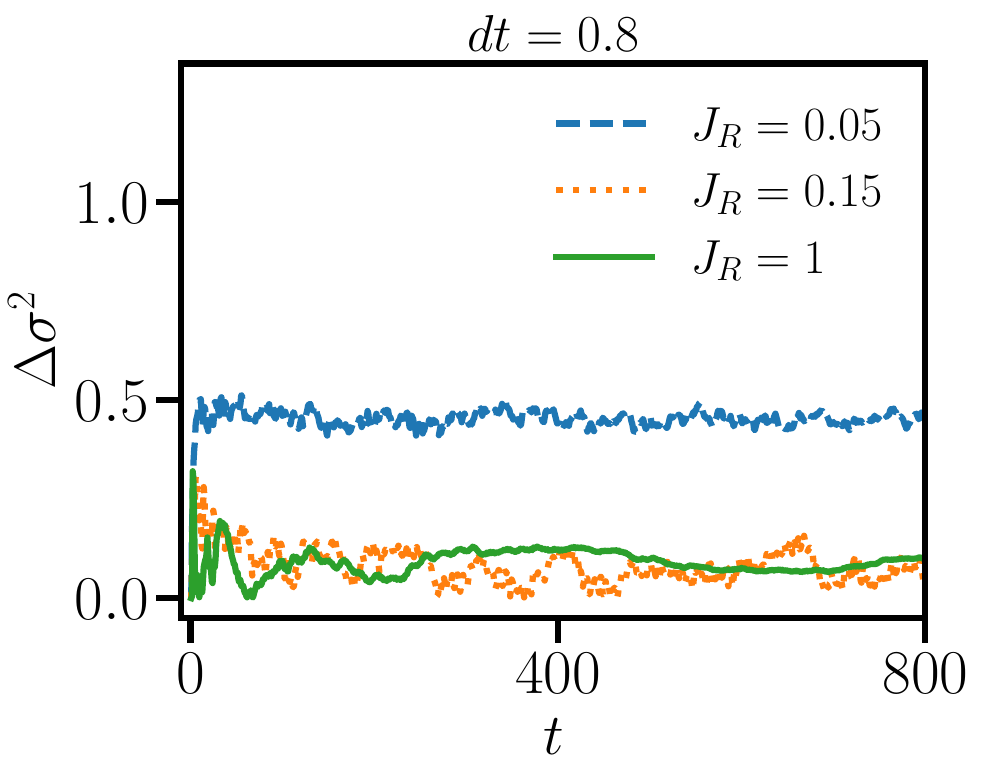}
     \caption{$\Delta \sigma^2$ vs. $t$ plots for three different $dt$ values. The dashed line represents $J_R=0.05$, the dotted line represents $J_R=0.15$, and the solid line represents $J_R=1$ (Hermitian case). Results are for system size $L=3001$, and $J_L=1$, $V=1$. }
     \label{sigma_diff}
\end{figure*}

\section{Effect of dephasing}
\label{Dephasing}
{\color{black}
So far, our study has established that the semiclassical versions of the non-Hermitian models (Model I and Model II) exhibit localization transitions whose critical points can be determined analytically. However, except for certain special choices of $\beta$, the semiclassical transition points do not coincide with the exact quantum transition points. In contrast, for the Hermitian models, the classical and quantum transition points coincide (see Appendix~\ref{app_A}).

This observation naturally raises the following question: does the discrepancy arise because quantum interference effects are substantially stronger in the non-Hermitian models than in their Hermitian counterparts? If so, the exact quantum transition would be intrinsically more "quantum" in nature and therefore beyond the predictive power of a semiclassical description.

To investigate this possibility, we study dephasing dynamics for the lattice realization of Model I, $H_I$, and compare the resulting dynamics with the exact quantum evolution. We begin with a pure initial state,
$|\psi(0)\rangle=\sum_j \psi_j |j\rangle$, 
where $|j\rangle$ denotes the lattice-site basis. The corresponding initial density matrix is 
$\rho(0)=|\psi(0)\rangle\langle\psi(0)|$.

For a chosen dephasing interval $dt$, the density matrix evolves according to
\[
\rho(t+dt)=
\frac{U(dt)\rho(t)U^\dagger(dt)}
{\mathrm{Tr}[U(dt)\rho(t)U^\dagger(dt)]},
\]
where $U(dt)=e^{-iH_I dt}$. After each evolution step, we completely remove the off-diagonal components of the density matrix,
\[
\rho_{m,n}(t+dt)\rightarrow
\rho^{d}_{m,n}(t+dt)\delta_{m,n},
\]
thereby replacing the evolved state by its fully dephased counterpart $\rho^d$. Repeating this procedure generates a dynamics in which quantum coherence is continuously suppressed. Since the dephased state contains only diagonal elements, it is generally a mixed state.

The rationale behind this construction is straightforward. For large dephasing intervals, $dt\rightarrow\infty$, the dephasing events become rare and the dynamics approaches the exact quantum evolution. In contrast, for small $dt$, coherence is removed more frequently, leading to a stronger modification of the dynamics. Consequently, if interference effects play a significant role in the exact dynamics, the dephased and exact evolutions should differ substantially for small $dt$. Conversely, if interference effects are weak, the two dynamics should remain similar even under frequent dephasing.

To quantify this effect, we calculate the second moment of the wave-packet propagation under the dephased dynamics,
\[
\sigma_d^2(t)=\sum_i i^2 \rho^{d}_{i,i}(t),
\]
and compare it with the corresponding quantity $\sigma^2(t)$ obtained from the exact quantum evolution. As a measure of the deviation between the two dynamics, we define
\[
\Delta\sigma^2=
\frac{|\sigma_d^2-\sigma^2|}{\sigma^2}.
\]

Figure~\ref{sigma_diff} presents $\Delta\sigma^2$ at $V=1$ for several values of the dephasing interval $dt$. For $dt=0.4$, we observe that the relative difference increases with decreasing the ratio $J_R/J_L$, which serves as a measure of the non-Hermiticity. Increasing the dephasing interval to $dt=0.6$, we find that $\Delta\sigma^2$ becomes nearly zero in the Hermitian limit, while a finite deviation persists for the non-Hermitian cases considered here. For $dt=0.8$, the data corresponding to $J_R=0.15$ and the Hermitian model are already very close to zero, whereas systems with stronger non-Hermiticity continue to exhibit a noticeable difference.

These observations suggest the existence of a characteristic dephasing timescale, $dt_c$, beyond which the dephased and exact dynamics become nearly indistinguishable. A small value of $dt_c$ implies that even frequent removal of quantum coherence has little effect on the dynamics, indicating relatively weak interference effects. Conversely, a larger $dt_c$ signifies that coherence must be preserved over longer timescales to accurately reproduce the exact dynamics, reflecting stronger quantum interference.

Our results indicate that $dt_c$ increases with increasing non-Hermiticity. This suggests that interference effects become progressively more important as Hermiticity is broken. Such enhanced interference may ultimately be responsible for the failure of the semiclassical description to accurately predict the localization transition in the non-Hermitian models.
}
\section{conclusion}\label{secVI}
In the Hermitian Aubry–André (AA) model, the delocalization–localization transition admits a clear semiclassical interpretation: the classical phase-space portrait develops separatrices precisely at the quantum critical point, and the semiclassical Husimi evolution correctly predicts the location of this transition~\cite{semiclassical_1}. In this work, we ask whether an analogous correspondence persists in non-Hermitian quasiperiodic lattice models \cite{non_hermitian_longhi_1,non_hermitian_longhi_2}, which are also known to exhibit localization–delocalization transitions.

To answer the proposed question, we analyze the Husimi distribution of the quantum state~\cite{Semiclassical_prl_1,semi_classical_prl_2}, which provides a phase-space representation of the underlying quantum dynamics. Starting from the continuum limit of the non-Hermitian lattice model, we compute the time evolution of the Husimi distribution for an initial coherent state and compare its propagation speed with that obtained directly from the lattice dynamics. The two descriptions show good qualitative agreement.
In contrast, the semiclassical Husimi evolution fails to reproduce the quantum localization–delocalization transition point. This failure is consistent with the fact that the trajectory equations governing the semiclassical Husimi dynamics do not predict the correct transition, as demonstrated by our fixed-point analysis, even though these trajectories exhibit a far richer and more intricate structure than their Hermitian counterparts. This highlights a fundamental limitation of semiclassical intuition in non-Hermitian quasiperiodic systems and points to qualitatively new mechanisms governing their localization behavior.
Furthermore, unlike the Hermitian Aubry–André case, the transition point of the non-Hermitian lattice model depends explicitly on the incommensurability parameter $\beta$, which characterizes the irrational quasiperiodicity of the potential. We identify a special value of $\beta$ for which the classical and quantum transition points coincide.

This naturally leads to a more fundamental question: are non-Hermitian systems more susceptible to a loss of classicalness than their Hermitian counterparts? 
\textcolor{black}{We compared our results with those obtained from dephased dynamics and found that the agreement deteriorates as the non-Hermiticity is increased, particularly near the quantum transition point. This observation suggests that quantum interference effects become increasingly important with increasing non-Hermiticity. Consequently, the growing role of interference may be responsible for the inability of the semi-classical description to accurately predict the location of the quantum transition point. }
Since the validity of semiclassical descriptions relies on the persistence of classical phase-space structures, an enhanced breakdown of classical correspondence could underlie the failure of semiclassical Husimi dynamics observed in non-Hermitian systems.
Moreover, the explicit $\beta$ dependence of the classically predicted localization transition points suggests that the incommensurability parameter plays a much more significant role in non-Hermitian settings, potentially influencing even the quantum localization transition. In Hermitian quasiperiodic models, $\beta$ is typically taken for granted, as the transition point is independent of its precise value. Our results indicate that this assumption may no longer hold once non-Hermiticity is introduced, and that the role of $\beta$ warrants careful re-examination in non-Hermitian quantum lattice models.

On the other hand, if one interprets this as a limitation of the semiclassical analysis—namely, its inability to accurately capture the localization transition for arbitrary values of $\beta$—our findings nevertheless reveal an important positive aspect. Specifically, we identify a nontrivial and experimentally relevant parameter regime (moderately large $\beta$) in which semiclassical correspondence remains quantitatively reliable within a finite time window in the non-Hermitian system, even though it fails in the corresponding Hermitian counterpart. 
Such a regime would not have been evident without constructing the explicit classical phase-space correspondence. Our analysis therefore demonstrates that classical Husimi evolution is not merely a qualitative diagnostic tool, but can provide an efficient and quantitatively accurate description of the underlying quantum dynamics over physically relevant timescales.

An intriguing direction is the construction of non-Hermitian models in which the quantum localization transition itself becomes explicitly $\beta$ dependent. Such a scenario would be especially appealing from an experimental standpoint, since $\beta$ can be tuned by varying the wavelength ratio of laser potentials in optical lattice realizations. This tunability would offer a direct and controlled way to explore distinct localization transition points within a single experimental platform.

%For completeness, we have also carried out a similar analysis for a Hermitian generalized Aubry–André model(GAA), which exhibits an energy-dependent critical parameter for the localization–delocalization transition at the quantum level. In contrast, the classical trajectory analysis yields a global, energy-independent critical value of the parameter, identified by the emergence of separatrices. 
%The evolution of classical trajectories does not provide a clear signature of a mobility edge. Nevertheless, for a given energy(saddle point energy), the quantum transition point coincides with the classical critical parameter. Similar to the non-Hermitian case, the speed of propagation extracted from the quantum Husimi and the lattice model shows almost a similar nature, while the velocity extracted from the semiclassical Husimi drops early at the critical parameter value predicted from the trajectory analysis.

\begin{acknowledgments}
R.M. acknowledges the DST-Inspire fellowship from the Department of Science and Technology, Government of India, SERB start-up grant (SRG/2021/002152). BPM acknowledges the PDF
grant of IOE of Banaras Hindu University,Varanasi for the year 2025-26. The authors gratefully acknowledge S.~Y.~Ali for valuable discussions during the early stages of this project.  
\end{acknowledgments}

\appendix

\begin{figure*}[t]
    \centering
    \includegraphics[width=0.32\textwidth]{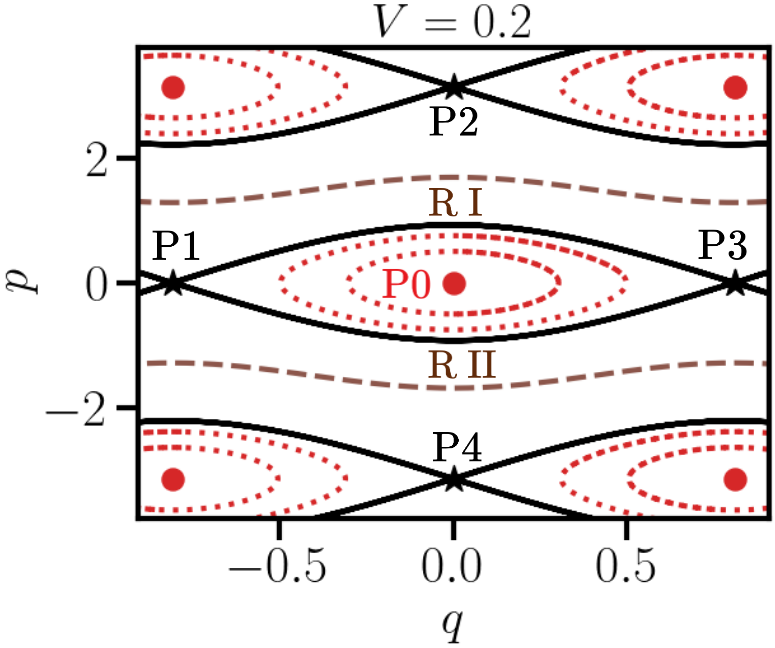}
    \includegraphics[width=0.32\textwidth]{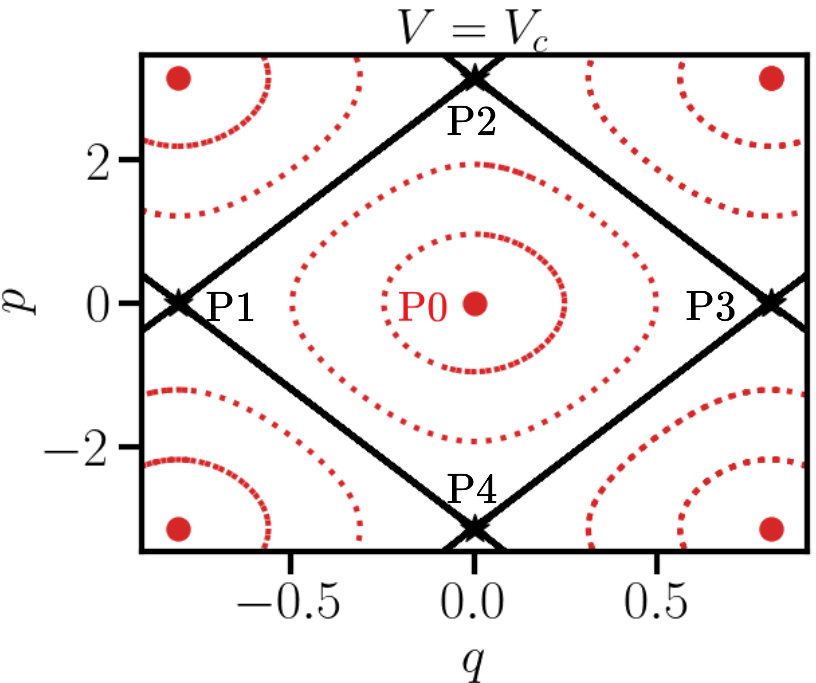}
    \includegraphics[width=0.32\textwidth]{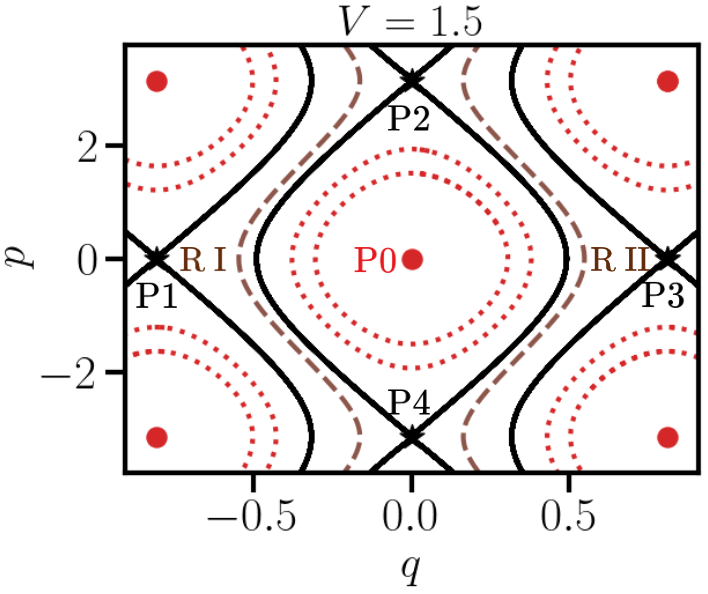}
    \caption{Trajectory evolution goverened by the semiclassical Hamiltonian [Eq. (\ref{semi_H}) at different $V$ values in the Hermitian case. Trajectory evolution shows a transition near $V_c=J_L=J_R$. Data are for $J_L=1, J_R=1$.  }
  \label{H_traj}
\end{figure*}

\begin{figure}[!h]
    \centering
    \includegraphics[width=0.45\textwidth]{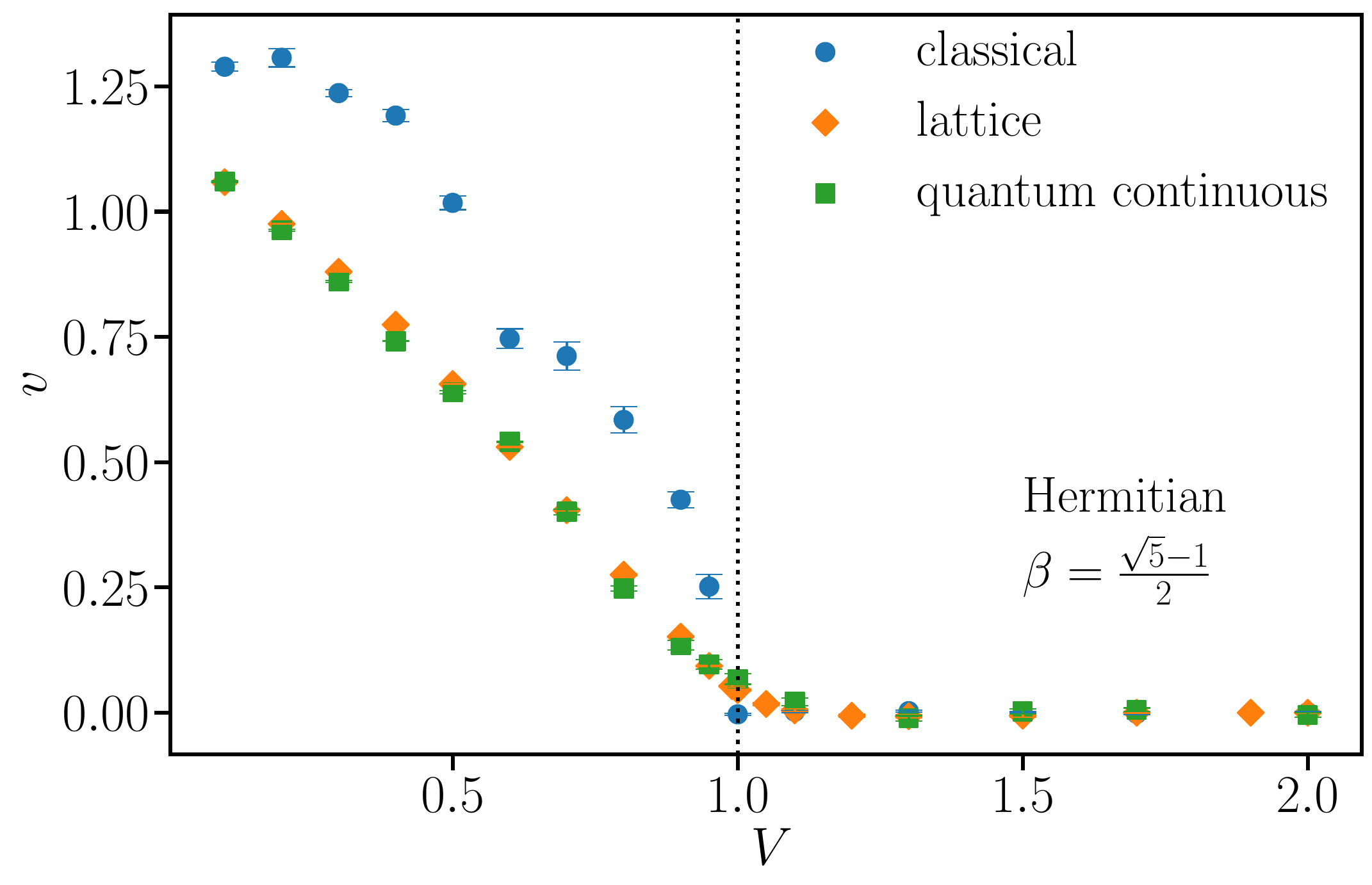}
    
    \caption{ Comparison of the velocity of the excitation of the Hermitian AA lattice Model with the excitation velocity extracted from the Husimi distribution of the quantum continuous Model, as well as the velocity extracted from the semiclassical Husimi evolution. \textcolor{black}{Results shown here are extrapolated thermodynamic findings}. }   
    \label{comp_Hermitian}
\end{figure}

\begin{figure*}
    \centering
    \includegraphics[width=0.85\textwidth, height=0.65\textwidth]{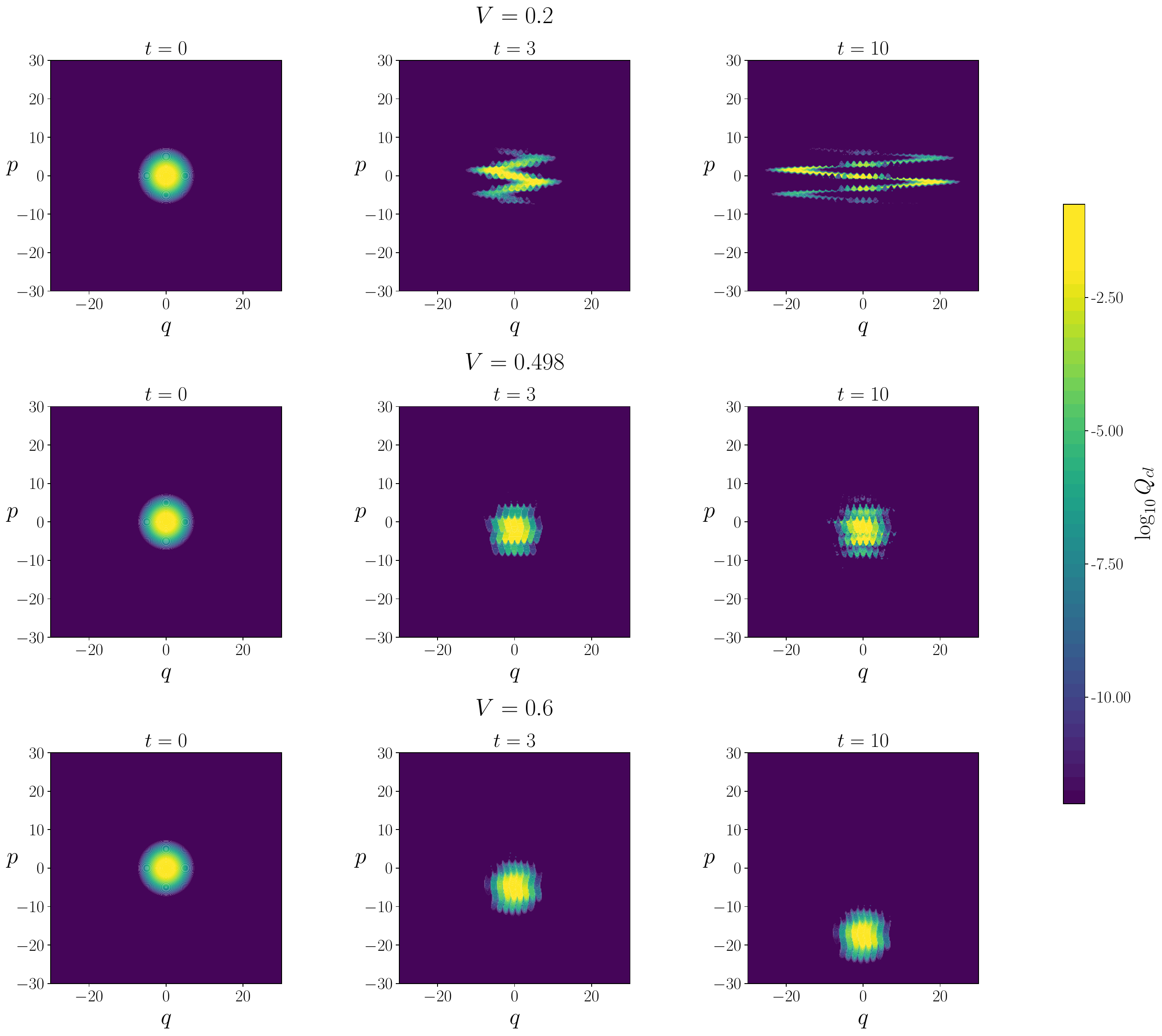}
   
    \caption{Semiclassical Husimi distribution at different time instances starting from coherent state $|z=0\rangle$ and evolving under the Hamiltonian [Eq. (\ref{semi_H_2})]. Data is for $V=0.2$, $0.498, 0.6$. The color code represents the Husimi values. }
    \label{semi_Husimi_1}
\end{figure*}

\section{Trajectory analysis and Husimi dynamics of Hermitian AA model}
\label{app_A}
For the sake of completeness, in this appendix we present the classical trajectory analysis of the Hermitian Aubry–André (AA) model~\cite{aubry.1980}. We find that the transition from the delocalized to the localized phase occurs at $V_c = J$ in both the quantum and classical versions of the model.  The lattice model is given by,
\begin{equation}
    H_{AA}=J\sum_i (  c_i^\dagger c_{i+1}+  c_{i+1}^\dagger c_i ) + 2V\sum_i  \cos(2\pi \beta i)c_i^{\dagger}c_i,
    \label{lattice_model}
\end{equation}\\
here, $c_i^\dagger, c_i$ are the creation and annihilation operators respectively, and $V$ is the strength of the quasiperiodic potential. $\beta$ is an irrational number. In our calculation we take $\beta=\frac{\sqrt{5}-1}{2}$.
The semiclassical version of the above Hamiltonian,  is given by~\cite{semiclassical_1, AA_semi_drive},
\begin{equation}
    H^{c}_{AA}=2J\cos p + 2V\cos(2\pi\beta q).
    \label{semi_H}
\end{equation}\\
In our calculation, we take $J=1$. The Hamiltonian equations of motion are,
\begin{eqnarray}
      &\dot{q}=-\frac{\partial  H^{c}_{AA}}{\partial p}=2\sin p\nonumber\\
    &\dot{p}=\frac{\partial  H^{c}_{AA}}{\partial q}=-4\pi\beta V \sin (2\pi\beta q).
    \label{AA_dyn_eq}
\end{eqnarray}
  For $V = 0$, the potential is absent, and all trajectories are spatially extended, exhibiting ballistic motion with the momentum remaining constant throughout the dynamics. In the opposite limit, $V \to \infty$, the potential term dominates: the kinetic energy becomes negligible, trajectories become trapped (localized) in space by the potential, and are correspondingly delocalized in momentum. In the intermediate regime, a crossover occurs, characterized by the emergence of separatrices that delineate three distinct types of trajectories.

%At $V=1$, the two separatrices merge together\cite{didov2020transport,semiclassical_1}. This is the critical value of the parameter $V$ above which
%the potential part dominates over the kinetic one, and all classical trajectories are localized in space by the potential barriers.
Below, we try to understand this from the fixed point(stationary points) analysis.

Once again, we follow the same strategy as discussed in the main text. First, we identify the fixed points of the dynamical equations as,
\begin{align*}
    &\dot{q}=0 ~~~\text{gives}~~~ p_0=n\pi, ~~~~n=0,\pm1,\pm 2\cdots.\\
    &\dot{p}=0 ~~~\text{gives}~~~ q_0=\frac{n\pi}{2\pi\beta}, ~~~~n=0,\pm1,\pm 2\cdots.
\end{align*}

The eigenvalues of the Jacobian matrix, obtained by linearizing Eq.~\eqref{AA_dyn_eq}, are 
\begin{align*}
    &(q_0,p_0)=(0,0)~~~\text{gives}~~~  \lambda=\pm 4Vi (2\pi\beta)^2 \\
    &(q_0,p_0)=(\pi/(2\pi\beta),\pi)~~~\text{gives}~~~\lambda=\pm 4Vi (2\pi\beta)^2  \\
    &(q_0,p_0)=(0,\pi)~~~\text{gives}~~~ \lambda=\pm 4V (2\pi\beta)^2\\
    &(q_0,p_0)=(\pi/(2\pi\beta),0)~~~\text{gives}~~~ \lambda=\pm 4V (2\pi\beta)^2.
\end{align*}

Hence, a slight departure from the stable fixed points $(0,0)$ and $(\pi/(2\pi\beta),\pi)$ leads to periodic, bounded orbits. In contrast, small deviations from the saddle points $(0,\pi)$ and $(\pi/(2\pi\beta),0)$ give rise to unbounded trajectories, as the corresponding linearized dynamics has one positive real eigenvalue.
Since the dynamical equations in Eq.~\ref{AA_dyn_eq} describe a time-independent, energy-conserving Hamiltonian system, each trajectory in the phase portrait corresponds to a fixed energy. In order to eliminate the separation between regions I and II for $V = 0.2$ (see Fig.~\ref{H_traj}), one must ensure that a single trajectory passes through multiple saddle points, for example P1 and P2, or P1 and P4, and so on. This is possible only when the energies associated with the corresponding saddle points coincide, i.e., when the saddle energies of points such as P2 and P3 merge. This implies, 
\begin{align*}
     &H^{c}_{AA}\big{|}_{0,\pi}=H^{c}_{AA}\big{|}_{\pi/{2\pi\beta},0},\\
    &-2+2V_c=2-2V_c.
\end{align*}
We get $V_c=1$.A similar argument can be drawn from a slope analysis, as discussed in the main text, where the dynamical equations are not Hamiltonian in nature. Note that the critical potential $V_c$ is independent of $\beta$, in contrast to the non-Hermitian case. These results can also be confirmed by performing a numerical trajectory analysis~\cite{didov2020transport,semiclassical_1}, as shown in Fig.~\ref{H_traj}.

Finally, in Fig.~\ref{comp_Hermitian}, we compare the excitation speed $v$ as a function of $V$ for the lattice model, the quantum Husimi calculation in the continuum model, and the classical Husimi distribution, following the same procedure as in the main text for the non-Hermitian case. We find that the phase transition point is correctly captured even by the classical Husimi dynamics, and, unlike the non-Hermitian model, no $\beta$ dependence is observed. \textcolor{black}{Furthermore, in the Hermitian case, the nature of transport at the transition point, $V=J$, differs significantly between the classical and quantum descriptions. For the lattice and quantum continuum models, the wave-packet width exhibits diffusive scaling, $\sigma \sim t^{1/2}$. In contrast, the semiclassical Husimi dynamics show a saturated behavior, $\sigma \sim t^{0}$, indicating the absence of spreading.
Nevertheless, according to the definition of the propagation velocity given in Eq.~\ref{Eq: def_v} of the main text, the velocity remains zero at the transition point for both the classical and quantum models. Thus, despite the qualitative differences in their transport properties at criticality, neither model supports ballistic propagation. This behavior is in sharp contrast to the non-Hermitian case discussed in the main text, where the quantum models retain a finite propagation velocity at the transition while the corresponding classical dynamics do not.
}

\begin{figure*}
    \centering
    \includegraphics[width=0.42\textwidth]{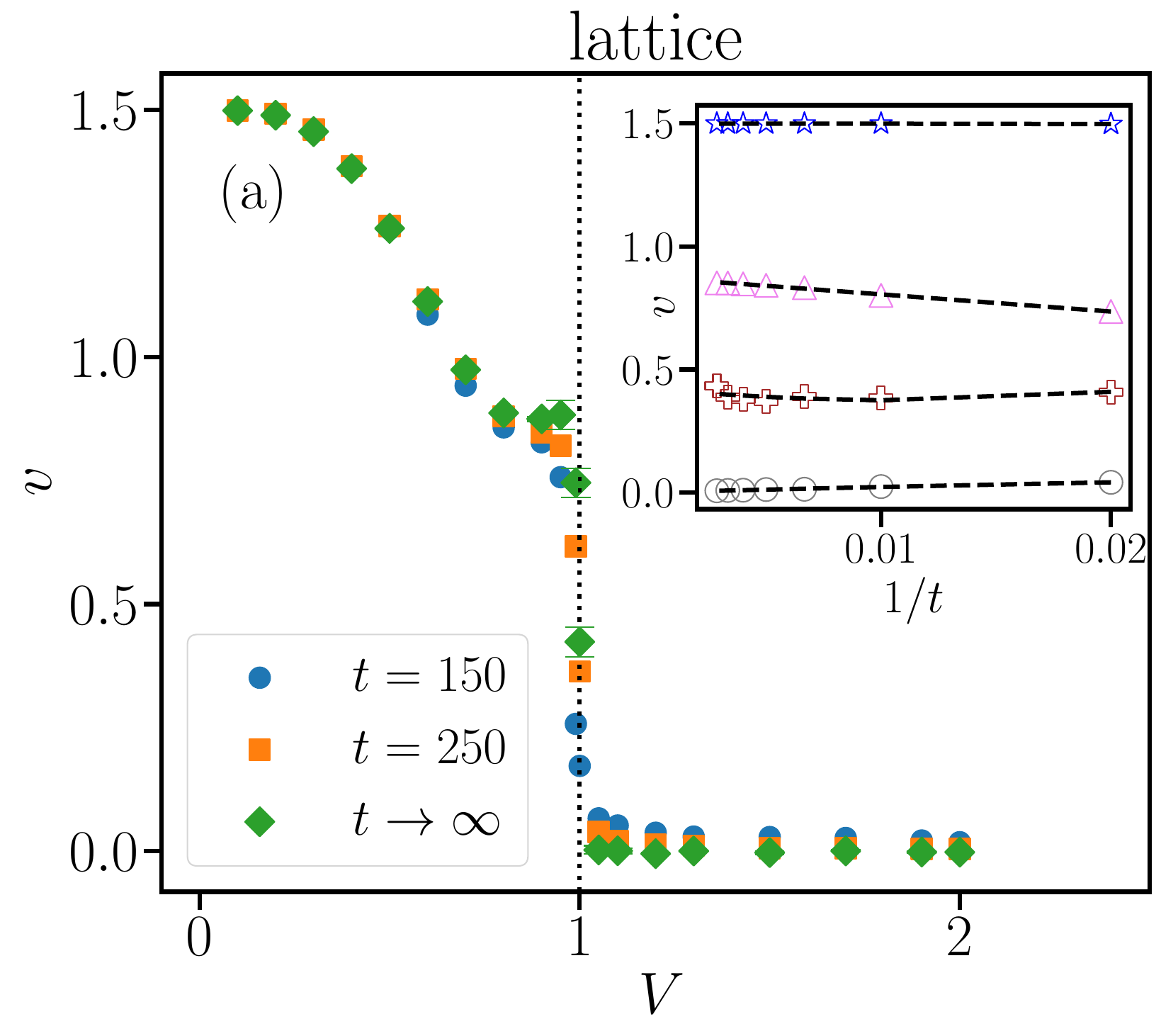}
\includegraphics[width=0.42\textwidth]{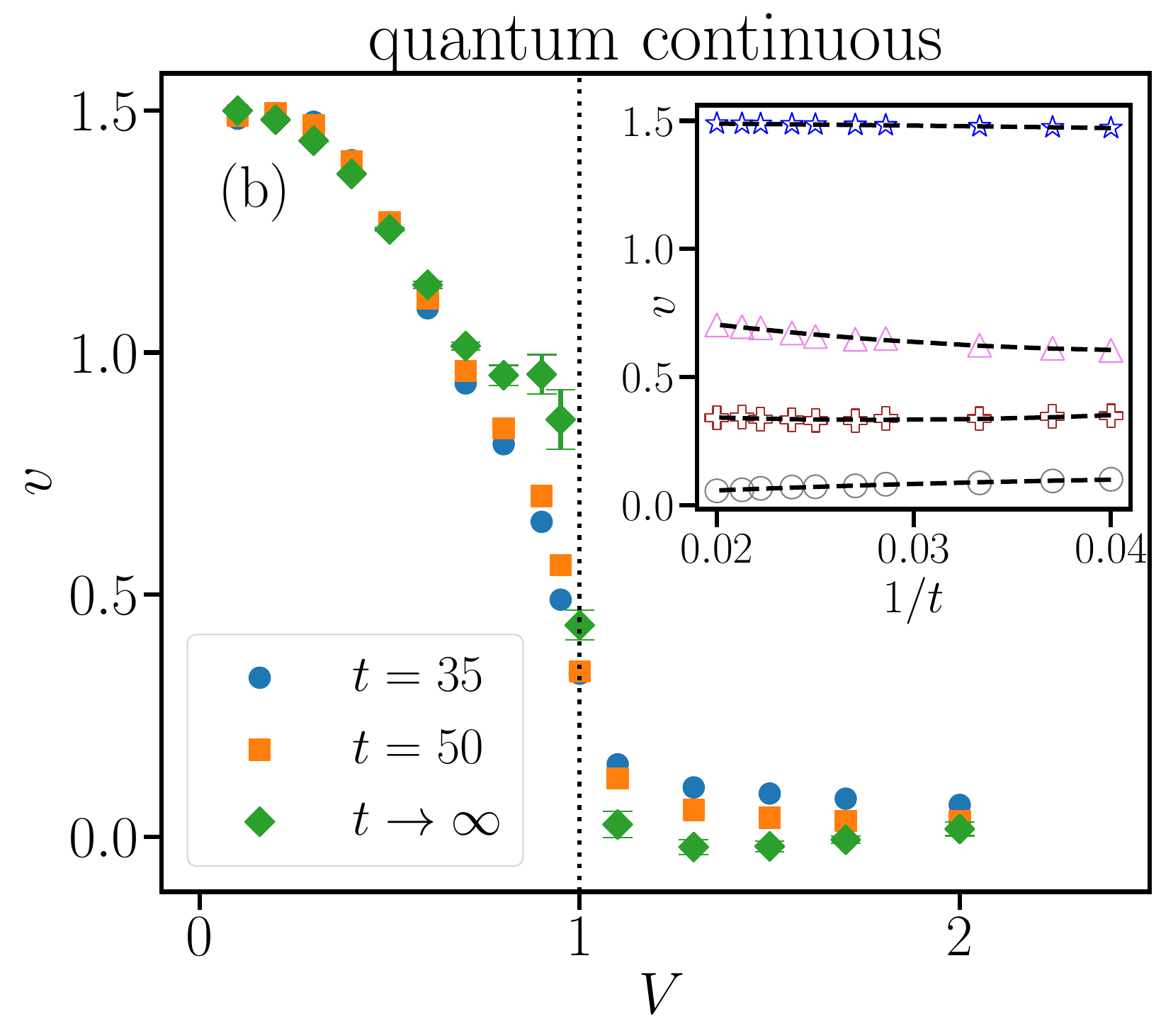}
\includegraphics[width=0.42\textwidth]{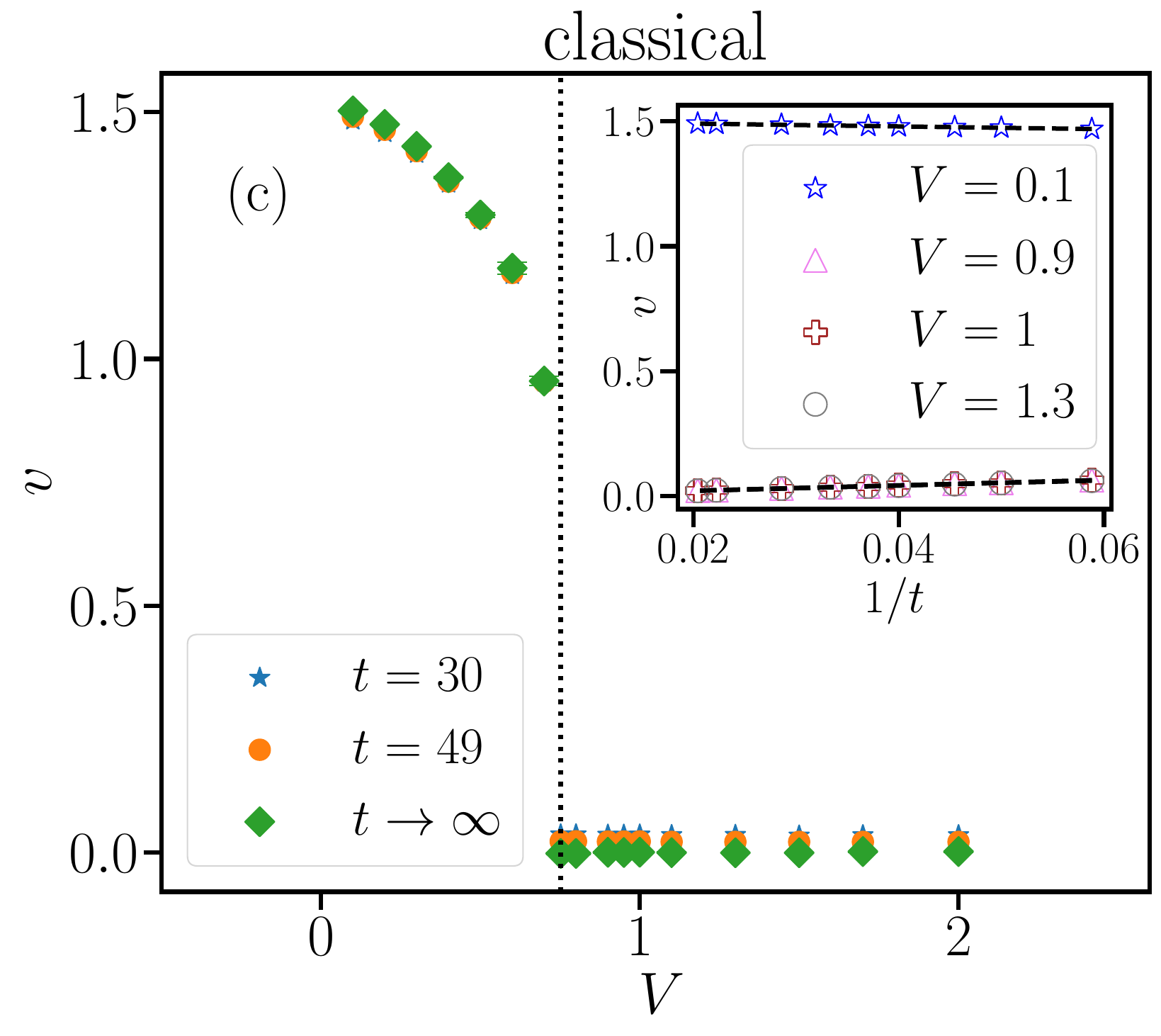}
\caption{$v$ vs. $V$ for different $t$ values. Green diamond points represent the extrapolated thermodynamic result. The inset shows the fitting of $v$ as a function of $1/t$ as $v= A+ \frac{B}{t} + \frac{C}{t^2}$. Data are for Model I. The (a), (b), and (c) correspond to lattice, quantum continuous, and semiclassical dynamics, respectively.}
\label{scaling}
\end{figure*}

\section{Semiclassical Husimi dynamics of Model II}
\label{Model_II_Husimi}
\textcolor{black}{In the main text, we compared the second moment of the wave-packet spreading for the quantum and classical versions of Model I and Model II with $\beta=(\sqrt{5}-1)/2$ (see Fig.~\ref{class_qunaum_comp}). While the semiclassical Husimi dynamics for the classical version of Model I were presented in the main text, here we show the corresponding Husimi dynamics for the classical version of Model II.
The semiclassical Husimi distributions are displayed in Fig.~\ref{semi_Husimi_1}. It is evident that the distribution ceases to propagate along the $q$ direction for $V \gtrsim 0.498$. This observation is consistent with our extrapolation analysis, which predicts the semiclassical transition point of Model II to be $V_c = 0.498$. Thus, the Husimi dynamics provide direct visual evidence of the localization transition in the classical counterpart of Model II.
}

\section{Thermodynamic scaling of the speed of excitation propagation $v$ }
 \label{extrapolated_results}
{\color{black}
As mentioned in the main text, the definition of the propagation velocity $v$ [Eq.~\eqref{Eq: def_v}] requires taking the limit $t\to\infty$ in order to accurately determine the localization transition. In practice, reaching this limit is challenging. For lattice models, one must simultaneously consider the thermodynamic limit $L\to\infty$; otherwise, the wave packet eventually reaches the system boundaries, preventing an accurate evaluation of $v$. For continuum models, although there is no underlying lattice, the Hamiltonian is represented in a harmonic-oscillator basis, and the accessible Hilbert-space dimension must be increased to faithfully capture the long-time dynamics. This rapidly increases the computational cost and makes simulations at very long times challenging.

To overcome these limitations, we employ an extrapolation procedure. We first simulate the dynamics over the longest accessible time window for which the extracted velocity $v(t)$ remains unaffected by finite-size effects in the lattice models or by Hilbert-space truncation effects in the continuum models. We then plot $v(t)$ as a function of $1/t$ and fit the data using the scaling form
$v(t)=A+\frac{B}{t}+\frac{C}{t^2}$.
The extrapolated propagation velocity is obtained from the fitting parameter $A$, corresponding to
$v=\lim_{t\to\infty}v(t)=A$,
which is precisely the quantity defined in Eq.~\eqref{Eq: def_v} of the main text.

Examples of this extrapolation procedure for the lattice, quantum continuum, and classical models are shown in Fig.~\ref{scaling}. The insets display the fits to the numerical data, while the main panels illustrate how the finite-time estimates gradually approach the extrapolated asymptotic value as $t$ increases.}

%In the main text, we have shown the extrapolated thermodynamic data ($t \to \infty$), as it is almost impossible to simulate the long-time dynamics for the quantum continuous models as Hilbert space dimension grows as $\sim |z|^2$, as we have mentioned in the main text. To get the long-time data, we have fitted the speed of propagation $v$ as a function of $1/t$ with a second-order polynomial function as 
%$v= A + B\frac{1}{t} + C \frac{1}{t^2}$, 
%up to the time we can go with the accessible Hilbert space dimension, and identify $A$ as the extrapolated thermodynamic value. In Fig.~\ref{scaling}, the inset shows the second-order polynomial fitting of $v$ as a function of $1/t$. For the three different dynamics(lattice, quantum continuous, and classical) in the left, middle, and right panels, respectively, for Model I. In the main panel, we compare the actual time data with the extrapolated thermodynamic data, and we see that as time increases, the actual time data approaches the extrapolated results. We check the convergence of the data for each time with the increasing system size(lattice) and Hilbert space dimension(quantum continuous) as well.   } 

\section{Variance of the propagation}
\begin{figure}[!h]
    \centering
    \includegraphics[width=0.45\textwidth]{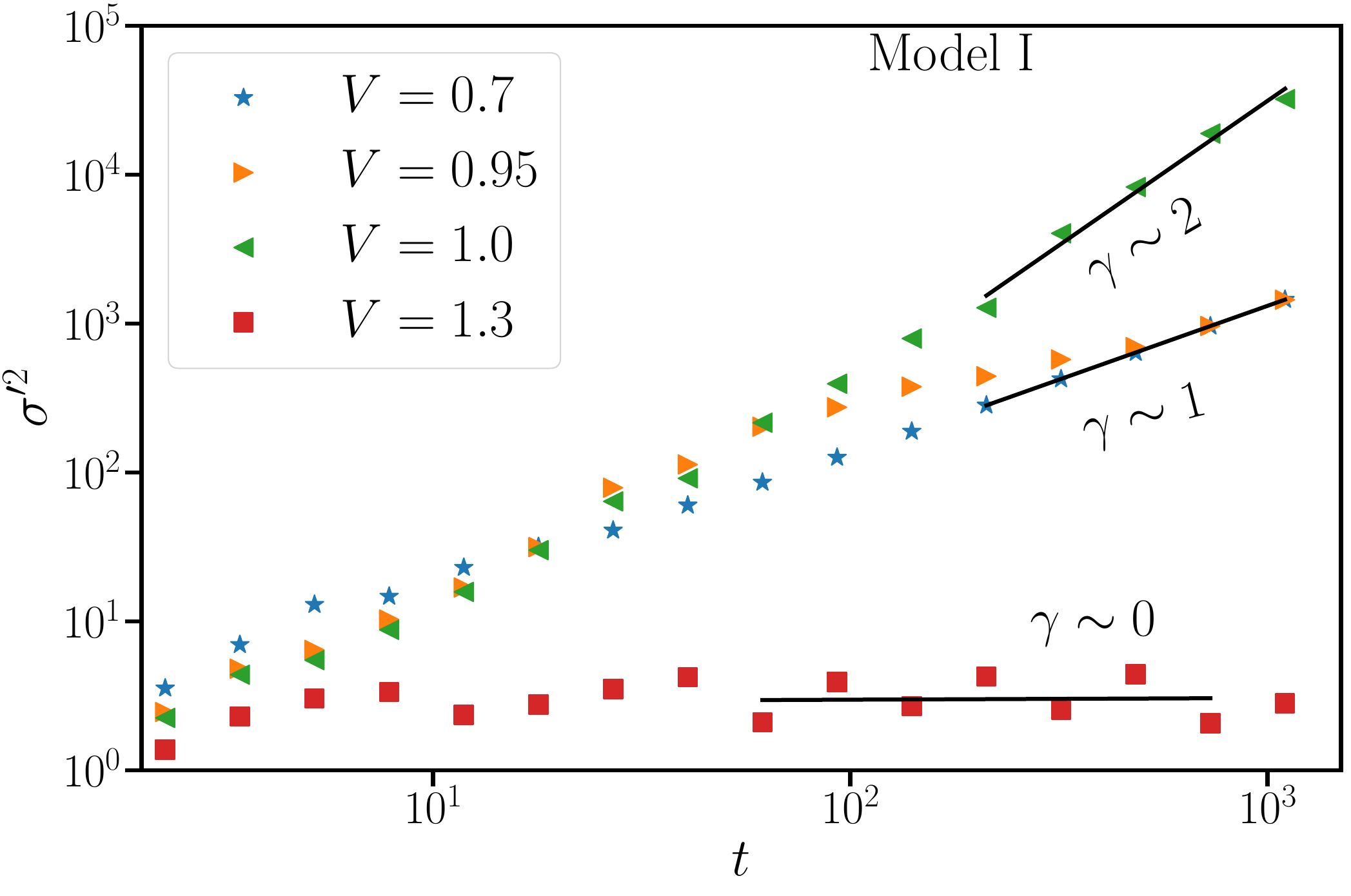}
    
    \caption{ $\sigma'^2$ vs. $t$ plot for the lattice Model I.}
  \label{actual_var}
\end{figure}
{\color{black}
In the main text, we used $\sigma^2(t)$, defined as the second moment of the wave-packet distribution, to characterize the dynamics. Since Model I exhibits unidirectional transport, the center of the wave packet itself drifts with time. Consequently, $\sigma^2(t)$ captures both the spreading of the wave packet and the motion of its center. Based on this quantity, we defined the propagation velocity, whose vanishing signals the onset of localization.

For completeness, we also compute the actual variance of the quantum lattice model, defined as
\[
\sigma'^2(t)=\frac{\sum_j j^2 |\psi_j(t)|^2}{\sum_j |\psi_j(t)|^2}
-\left[\frac{\sum_j j |\psi_j(t)|^2}{\sum_j |\psi_j(t)|^2}\right]^2,
\]
which explicitly removes the contribution from the drift. Thus, $\sigma'^2(t)$ measures only the spreading of the wave packet about its instantaneous center of mass. For Model II and the Hermitian case, where the spreading is symmetric, the center of mass remains stationary, and therefore $\sigma^2(t)$ and $\sigma'^2(t)$ are equivalent.

Interestingly, Fig.~\ref{actual_var} reveals a distinct behavior for Model I. For $V<V_c$, the variance scales as $\sigma'^2(t)\sim t$, indicating diffusive spreading around the moving center. At the critical point, $V=V_c$, we find $\sigma'^2(t)\sim t^2$ over a substantial time window, implying ballistic spreading. This suggests that the critical point is special: the wave packet exhibits  diffusive drift($\langle x\rangle\sim t$) and ballistic spreading. In contrast, throughout the extended phase ($V<V_c$), the drift remains diffusive—leading to the overall scaling $\sigma^2(t)\sim t^2$—while the spreading about the center is only diffusive, as captured by $\sigma'^2(t)$.}

{\color{black}
The observation that $\sigma'^2(t)$ exhibits diffusive scaling while $\sigma^2(t)$ remains ballistic for $V<V_c$ in Model I is markedly different from the Hermitian case, where both quantities display ballistic behavior. 
To gain analytical insight into this phenomenon, we consider the exactly solvable limit $V=0$. If the same behavior can be demonstrated in this limit, it would provide strong support for our numerical observations at finite $V$. In the absence of $V$, the Hamiltonian simplifies considerably and can be written as
\begin{align*}
    H_I(V=0)&= \sum_p \left[(J_R + J_L)\cos p + i (J_R-J_L)\sin p  \right] c_p^\dagger c_p ,\\
    &= \sum_p \left[\epsilon_p + i \eta_p  \right] c_p^\dagger c_p,
\end{align*}
with $p \in [0, 2\pi]$, $\epsilon_p=(J_R + J_L)\cos p$, and $\eta_p=(J_R-J_L)\sin p$. Starting from an initial state centred at the centre of the lattice, the time-evolved wavefunction could be written in the momentum basis as,
\begin{align*}
    |\psi(t)\rangle=\frac{1}{\sqrt{L}}\sum_p e^{-i\epsilon_p t + \eta_p t} |p\rangle.
\end{align*}
It is easy to show that the mean of the probability distribution associated with the above wave function evolves as,
\begin{align}
    \langle x(t)\rangle=\frac{t\int_0^{2\pi}  \partial_p \epsilon_p~e^{2 \eta_p t}  dp}{\int_0^{2\pi} e^{2 \eta_p t }dp}=-\frac{t(J_L+J_R)\int_0^{2\pi} \sin p~ e^{2 \eta_p t} dp}{\int_0^{2\pi} e^{2 \eta_p t }dp}.
   \label{mean_eq}
\end{align}
Similarly, the second moment $\sigma^2(t)=\langle x^2 (t)\rangle$ could be written as,
\begin{align}
\langle x^2 (t) \rangle&=2J_LJ_R t^2-\frac{(J_R^2 + J_L^2)t^2\int_0^{2\pi} \cos2p~ e^{2\eta_p t }dp}{\int_0^{2\pi} e^{2 \eta_p t} dp}\nonumber \\
    &~~~~~~~~~~~~~~~~+\frac{(J_R - J_L)t\int_0^{2\pi} \sin p~ e^{2\eta_p t }dp}{\int_0^{2\pi} e^{2 \eta_p t} dp}. \nonumber\\
    \label{sigma_eq}
\end{align}
\subsubsection{Hermitian case ($J_L=J_R=J$)}
In the Hermitian case ($J_L=J_R$), $\eta_p=0$, and therefore the integral in Eq.~\eqref{mean_eq} vanishes over a complete period, yielding $\langle x(t)\rangle=0$.

Similarly, as $\eta_p=0$, the second and third integral of Eq.~\eqref{sigma_eq} vanishes over a complete period and we get,
$\sigma^2(t)=\sigma'^2(t)=2J^2t^2$, where, $J_L=J_R=J$.
Hence, for the Hermitian model the growth of the second moment and variance with time is identical. A similar conclusion holds for Model II, with $V=0$.

\subsubsection{Non-Hermitian case ($J_L \ne J_R$)}
In the non-Hermitian case, where $J_L \ne J_R$, mean of the distribution can be calculated using Eq.~\eqref{mean_eq} and we get,
\begin{align*}
    \langle x(t)\rangle=(J_L+J_R)t\frac{I_1(a)}{I_0(a)}
\end{align*}
where $I_n(.)$ is the modified Bessel function of first kind~\cite{besel}, and $a=2(J_L-J_R)t$.\\

Similarly $\sigma^2$ can be calculated from the Eq.~\eqref{sigma_eq} as,
\begin{align*}
    \sigma^2(t)=2J_LJ_Rt^2 + (J_R^2+J_L^2)t^2\frac{I_2(a)}{I_0(a)}+(J_L-J_R)t \frac{I_1(a)}{I_0(a)}.
\end{align*}
In the asymptotic limit ($a\gg 0)$, 
\begin{align*}
    \frac{I_1(a)}{I_0(a)} \sim 1, ~~~\frac{I_2(a)}{I_0(a)} \sim 1,\text{ hence, and one gets,}
\end{align*}
\begin{align*}
    &\sigma^2(t)\simeq\langle x(t)\rangle^2 + (J_L-J_R)t, \\
     &\sigma^2(t)-\langle x(t)\rangle^2=(J_L-J_R)t,\\
     & \sigma'^2(t) = (J_L-J_R)t.
     \end{align*}
This result suggests that, in the asymptotic limit, the mean of the distribution propagates diffusively, while propagation of the variance $\sigma'^2\sim t$ is also diffusive,  which is in agreement with the results shown in Fig.~\ref{actual_var} for $V<V_c$.}
\bibliography{levy}
\end{document}